%%%%%%%%%%%%%%%%%%%%%
%  lanlmac vs harvmac
%%%%%%%%%%%%%%%%%%%%%
\openin 1 lanlmac
\ifeof 1
  \message{[Load harvmac]}
  \input harvmac
\else
  \message{[Load lanlmac]}
  \input lanlmac
\fi
\closein 1

%%%%%%%%%%%%%%%%%%%%%%
%  packages
%%%%%%%%%%%%%%%%%%%%%%
\input amssym
\input xymatrix
\input xyarrow
%\input epsf
%\input tables
%\draftmode
\noblackbox

%%%%%%%%%%%%%%%%%%%%%%
%  hypertex
%%%%%%%%%%%%%%%%%%%%%%
\newif\ifhypertex
\ifx\hyperdef\UnDeFiNeD
    \hypertexfalse
    \message{[HYPERTEX MODE OFF]}
    \def\hyperref#1#2#3#4{#4}
    \def\hyperdef#1#2#3#4{#4}
    \def\hypernoname{}
    \def\e@tf@ur#1{}
    
\else
    \hypertextrue
    \message{[HYPERTEX MODE ON]}
    
\fi

%%%%%%%%%%%%%%%%%%%%%%
%  Sizes
%%%%%%%%%%%%%%%%%%%%%%

\baselineskip=16pt plus 2pt minus 1pt
\parskip=2pt plus 16pt minus 1pt

%%%%%%%%%%%%%%%%%%%%%%
%  Tables
%%%%%%%%%%%%%%%%%%%%%%
\newcount\tabno
\tabno=1
\def\ltab#1{%
\let\labelflag=#1%
\def\numb@rone{#1}%
\ifx\labelflag\UnDeFiNeD{%
  \xdef#1{\the\tabno}%
  \writedef{#1\leftbracket{\the\tabno}}%
  \global\advance\tabno by1%
}%
\fi%
{\hyperref{}{table}{{\numb@rone}}{Table~{\numb@rone}}}}
\def\tabinsert#1#2#3{%
\let\flag=#1
\ifx\flag\UnDeFiNeD
  {\xdef#1{\the\tabno}
   \writedef{#1\leftbracket{\the\tabno}}
   \global\advance\tabno by1 }
\fi
\vbox{\bigskip #3 \smallskip
\leftskip 4pc \rightskip 4pc
\noindent\ninepoint\sl \baselineskip=11pt
{\bf{\hyperdef\hypernoname{table}{{#1}}{Table~{#1}}}.~}#2
\smallskip}
\bigskip}

%%%%%%%%%%%%%%%%%%%%%%
%  macros
%%%%%%%%%%%%%%%%%%%%%%
\def\ie{{\it i.e.\ }}
\def\eg{{\it e.g.\ }}
\def\cf{{\it cf.\ }}
% 'double' letters
\def\IC{{\Bbb C}}
\def\IP{{\Bbb P}}
\def\IQ{{\Bbb Q}}

\def\IW{{\Bbb W}}
\def\IZ{{\Bbb Z}}
% calligraphic letters

\def\cF{{\cal F}}

\def\cL{{\cal L}}

\def\uu{\mathop{\sl u}}
\def\vv{\mathop{\sl v}}
\def\var1{z}
%%%%%%%%%%%%%%%%%%%%%%
\def\rel#1{\underline{#1}}
\def\HGF#1#2{{{}_#1F_#2}}

%\def\HJ#1{{\bf HJ: #1}}
%\def\MS#1{{\bf MS: #1}}

%%%%%%%%%%%%%%%%%%%%%%
%  references
%%%%%%%%%%%%%%%%%%%%%%

%\AganagicDB
\lref\AganagicDB{
M.~Aganagic, A.~Klemm, M.~Mari\~no and C.~Vafa,
``The topological vertex,''
Commun.\ Math.\ Phys.\  {\bf 254}, 425 (2005)
[arXiv:hep-th/0305132].
%%CITATION = CMPHA,254,425;%%
}

%\AganagicGS
\lref\AganagicGS{
M.~Aganagic and C.~Vafa,
``Mirror symmetry, D-branes and counting holomorphic discs,''
arXiv:hep-th/0012041.
%%CITATION = HEP-TH/0012041;%%
}

%\AganagicNX
\lref\AganagicNX{
M.~Aganagic, A.~Klemm and C.~Vafa,
``Disk instantons, mirror symmetry and the duality web,''
Z.\ Naturforsch.\  A {\bf 57}, 1 (2002)
[arXiv:hep-th/0105045].
%%CITATION = ZNTFA,A57,1;%%
}

%\AganagicWQ
\lref\AganagicWQ{
M.~Aganagic, V.~Bouchard and A.~Klemm,
``Topological Strings and (Almost) Modular Forms,''
Commun.\ Math.\ Phys.\  {\bf 277}, 771 (2008)
[arXiv:hep-th/0607100].
%%CITATION = CMPHA,277,771;%%
}

%\AntoniadisZE
\lref\AntoniadisZE{
I.~Antoniadis, E.~Gava, K.~S.~Narain and T.~R.~Taylor,
``Topological amplitudes in string theory,''
Nucl.\ Phys.\  B {\bf 413}, 162 (1994)
[arXiv:hep-th/9307158].
%%CITATION = NUPHA,B413,162;%%
}

%\AspinwallJR
\lref\AspinwallJR{
P.~S.~Aspinwall,
``D-branes on Calabi-Yau manifolds,''
arXiv:hep-th/0403166.
%%CITATION = HEP-TH/0403166;%%
}

%\AspinwallPU
\lref\AspinwallPU{
P.~S.~Aspinwall and A.~E.~Lawrence,
``Derived categories and zero-brane stability,''
JHEP {\bf 0108}, 004 (2001)
[arXiv:hep-th/0104147].
%%CITATION = JHEPA,0108,004;%%
}

%\BayerAA
\lref\BayerAA{
A.~Bayer and C.~Cadman,
``Quantum cohomology of $[\IC^n/\mu_r]$,"
arXiv:0705.2160 [math.AG].
}

%\BershadskyCX
\lref\BershadskyCX{
M.~Bershadsky, S.~Cecotti, H.~Ooguri and C.~Vafa,
``Kodaira-Spencer theory of gravity and exact results for quantum string
amplitudes,''
Commun.\ Math.\ Phys.\  {\bf 165}, 311 (1994)
[arXiv:hep-th/9309140].
%%CITATION = CMPHA,165,311;%%
}

%BuchardAA
\lref\BouchardAA{
V.~Bouchard, and R.~Cavalieri,
``On the mathematics and physics of high genus invariants of $\IC^3/\IZ_3$,"
arXiv:0709.1453 [math.AG].
} 

%\BouchardYS
\lref\BouchardYS{
V.~Bouchard, A.~Klemm, M.~Mari\~no and S.~Pasquetti,
``Remodeling the B-model,''
arXiv:0709.1453 [hep-th].
%%CITATION = ARXIV:0709.1453;%%
}

%\BouchardGU
\lref\BouchardGU{
V.~Bouchard, A.~Klemm, M.~Mari\~no and S.~Pasquetti,
``Topological open strings on orbifolds,''
arXiv:0807.0597 [hep-th].
%%CITATION = ARXIV:0807.0597;%%
}

%\BriniRH
\lref\BriniRH{
A.~Brini and A.~Tanzini,
``Exact results for topological strings on resolved Y(p,q) singularities,''
arXiv:0804.2598 [hep-th].
%%CITATION = ARXIV:0804.2598;%%
}

%\CachazoJY
\lref\CachazoJY{
F.~Cachazo, K.~A.~Intriligator and C.~Vafa,
``A large N duality via a geometric transition,''
Nucl.\ Phys.\  B {\bf 603}, 3 (2001)
[arXiv:hep-th/0103067].
%%CITATION = NUPHA,B603,3;%%
}

%\CandelasDM
\lref\CandelasDM{
  P.~Candelas, X.~De La Ossa, A.~Font, S.~H.~Katz and D.~R.~Morrison,
  ``Mirror symmetry for two parameter models. I,''
  Nucl.\ Phys.\  B {\bf 416}, 481 (1994)
  [arXiv:hep-th/9308083].
  %%CITATION = NUPHA,B416,481;%%
}

%\CandelasHW
\lref\CandelasHW{
P.~Candelas, A.~Font, S.~H.~Katz and D.~R.~Morrison,
``Mirror symmetry for two parameter models. 2,''
Nucl.\ Phys.\  B {\bf 429}, 626 (1994)
[arXiv:hep-th/9403187].
%%CITATION = NUPHA,B429,626;%%
}

%\CandelasRM
\lref\CandelasRM{
P.~Candelas, X.~C.~De La Ossa, P.~S.~Green and L.~Parkes,
``A pair of Calabi-Yau manifolds as an exactly soluble superconformal
theory,''
Nucl.\ Phys.\  B {\bf 359}, 21 (1991).
%%CITATION = NUPHA,B359,21;%%
}

%\CandelasSE
\lref\CandelasSE{
P.~Candelas,
``Yukawa Couplings Between (2,1) Forms,''
Nucl.\ Phys.\  B {\bf 298}, 458 (1988).
%%CITATION = NUPHA,B298,458;%%
}

%\CecottiME
\lref\CecottiME{
S.~Cecotti and C.~Vafa,
``Topological antitopological fusion,''
Nucl.\ Phys.\  B {\bf 367}, 359 (1991).
%%CITATION = NUPHA,B367,359;%%
}

%\CoatesAA
\lref\CoatesAA{
T.~Coates, A.~Corti, H.~Iritani and H.-H.~Tseng
``Computing Genus-Zero Twisted Gromov-Witten Invariants,"
arXiv:math/0702234v3 [math.AG].
}

%\DeligneII
\lref\DeligneII{
P.~Deligne,
``Th\'eorie de Hodge~II,"
Publ.\ Math.\ I.H.E.S.\ {\bf 40}, 5 (1971).
}

%\DeligneIII
\lref\DeligneIII{
P.~Deligne,
``Th\'eorie de Hodge~III,"
Publ.\ Math.\ I.H.E.S.\ {\bf 55}, 5 (1974).
}

%\DijkgraafDJ
\lref\DijkgraafDJ{
R.~Dijkgraaf, H.~L.~Verlinde and E.~P.~Verlinde,
``Topological Strings In D < 1,''
Nucl.\ Phys.\  B {\bf 352}, 59 (1991).
%%CITATION = NUPHA,B352,59;%%
}

%\DouglasGI
\lref\DouglasGI{
M.~R.~Douglas,
``D-branes, categories and N = 1 supersymmetry,''
J.\ Math.\ Phys.\  {\bf 42}, 2818 (2001)
[arXiv:hep-th/0011017].
%%CITATION = JMAPA,42,2818;%%
}

%\GreeneUD
\lref\GreeneUD{
B.~R.~Greene and M.~R.~Plesser,
``Duality in Calabi-Yau moduli space,''
Nucl.\ Phys.\  B {\bf 338}, 15 (1990).
%%CITATION = NUPHA,B338,15;%%
}

%\GriffithsAA
\lref\GriffithsAA{
P.~Griffiths,
``On the periods of certain rational integrals: I,"
Ann.\ Math.\ {\bf 90}, 460 (1969).
}

%\GriffithsBB
\lref\GriffithsBB{
P.~Griffiths,
``A theorem concerning the differential equations satisfied by
normal functions associated to algebraic cycles,"
Am.\ J.\ Math.\ {\bf 101}, 94 (1979).
}

%\GukovYA
\lref\GukovYA{
S.~Gukov, C.~Vafa and E.~Witten,
``CFT's from Calabi-Yau four-folds,''
Nucl.\ Phys.\  B {\bf 584}, 69 (2000)
[Erratum-ibid.\  B {\bf 608}, 477 (2001)]
[arXiv:hep-th/9906070].
%%CITATION = NUPHA,B584,69;%%
}

%\HoriAA
\lref\HoriAA{
K.~Hori, S.~Katz, A.~Klemm, R.~Pandharipande, R.~Thomas, C.~Vafa, R.~Vakil and E. Zaslow,
``Mirror Symmetry,"
American Mathematical Society, 2003.
}

%\HoriJA
\lref\HoriJA{
K.~Hori and J.~Walcher,
``F-term equations near Gepner points,''
JHEP {\bf 0501}, 008 (2005)
[arXiv:hep-th/0404196].
%%CITATION = JHEPA,0501,008;%%
}

%\IntriligatorUA
\lref\IntriligatorUA{
K.~A.~Intriligator and C.~Vafa,
``Landau-Ginzburg Orbifolds,''
Nucl.\ Phys.\  B {\bf 339}, 95 (1990).
%%CITATION = NUPHA,B339,95;%%
}

%\KachruAN
\lref\KachruAN{
S.~Kachru, S.~H.~Katz, A.~E.~Lawrence and J.~McGreevy,
``Mirror symmetry for open strings,''
Phys.\ Rev.\  D {\bf 62}, 126005 (2000)
[arXiv:hep-th/0006047].
%%CITATION = PHRVA,D62,126005;%%
}

%\KachruIH
\lref\KachruIH{
S.~Kachru, S.~H.~Katz, A.~E.~Lawrence and J.~McGreevy,
``Open string instantons and superpotentials,''
Phys.\ Rev.\  D {\bf 62}, 026001 (2000)
[arXiv:hep-th/9912151].
%%CITATION = PHRVA,D62,026001;%%
}

%KaroubiBook
\lref\KaroubiBook{
M.~Karoubi and C.~Leruste,
``Algebraic topology via differential geometry,"
Cambridge University Press, Cambridge 1987.
}

%\KlemmBJ
\lref\KlemmBJ{
A.~Klemm, W.~Lerche, P.~Mayr, C.~Vafa and N.~P.~Warner,
``Self-Dual Strings and N=2 Supersymmetric Field Theory,''
Nucl.\ Phys.\  B {\bf 477}, 746 (1996)
[arXiv:hep-th/9604034].
%%CITATION = NUPHA,B477,746;%%
}

%\KlemmTX
\lref\KlemmTX{
A.~Klemm and S.~Theisen,
``Considerations of one modulus Calabi-Yau compactifications: Picard-Fuchs
equations, Kahler potentials and mirror maps,''
Nucl.\ Phys.\  B {\bf 389}, 153 (1993)
[arXiv:hep-th/9205041].
%%CITATION = NUPHA,B389,153;%%
}

%\KnappUW
\lref\KnappUW{
J.~Knapp and E.~Scheidegger,
``Towards Open String Mirror Symmetry for One-Parameter Calabi-Yau
Hypersurfaces,''
arXiv:0805.1013 [hep-th].
%%CITATION = ARXIV:0805.1013;%%
}

%\KontsevichAA
\lref\KontsevichAA{
M.~Kontsevich,
``Homological Algebra of Mirror Symmetry,"
arXiv:alg-geom/9411018.
}

%\KreflSJ
\lref\KreflSJ{
D.~Krefl and J.~Walcher,
``Real Mirror Symmetry for One-parameter Hypersurfaces,''
arXiv:0805.0792 [hep-th].
%%CITATION = ARXIV:0805.0792;%%
}

%\LabastidaZP
\lref\LabastidaZP{
J.~M.~F.~Labastida and M.~Mari\~no,
``Polynomial invariants for torus knots and topological strings,''
Commun.\ Math.\ Phys.\  {\bf 217}, 423 (2001)
[arXiv:hep-th/0004196].
%%CITATION = CMPHA,217,423;%%
}

%\LazaroiuJM
\lref\LazaroiuJM{
C.~I.~Lazaroiu,
``Generalized complexes and string field theory,''
JHEP {\bf 0106}, 052 (2001)
[arXiv:hep-th/0102122].
%%CITATION = JHEPA,0106,052;%%
}

%\LazaroiuMD
\lref\LazaroiuMD{
C.~I.~Lazaroiu,
``D-brane categories,''
Int.\ J.\ Mod.\ Phys.\  A {\bf 18}, 5299 (2003)
[arXiv:hep-th/0305095].
%%CITATION = IMPAE,A18,5299;%%
}

%\LercheCK
\lref\LercheCK{
W.~Lerche, P.~Mayr and N.~Warner,
``Holomorphic N = 1 special geometry of open-closed type II strings,''
arXiv:hep-th/0207259.
%%CITATION = HEP-TH/0207259;%%
}

%\LercheUY
\lref\LercheUY{
W.~Lerche, C.~Vafa and N.~P.~Warner,
``Chiral Rings in N=2 Superconformal Theories,''
Nucl.\ Phys.\  B {\bf 324}, 427 (1989).
%%CITATION = NUPHA,B324,427;%%
}

%\LercheWM
\lref\LercheWM{
W.~Lerche, D.~J.~Smit and N.~P.~Warner,
``Differential equations for periods and flat coordinates in two-dimensional
topological matter theories,''
Nucl.\ Phys.\  B {\bf 372}, 87 (1992)
[arXiv:hep-th/9108013].
%%CITATION = NUPHA,B372,87;%%
}

%\LercheYW
\lref\LercheYW{
W.~Lerche, P.~Mayr and N.~Warner,
``N = 1 special geometry, mixed Hodge variations and toric geometry,''
arXiv:hep-th/0208039.
%%CITATION = HEP-TH/0208039;%%
}

%\LibgoberAA
\lref\LibgoberAA{
A.~Libgober and J.~Teitelbaum.
``Lines on Calabi-Yau complete intersections, mirror symmetry, and
Picard Fuchs equations,"
In.\ Math.\ Res.\ Not.\ {\bf 1993}, 13 (1993)
[arXiv:alg-geom/9301001].
}

%\MayrXK
\lref\MayrXK{
P.~Mayr,
``N = 1 mirror symmetry and open/closed string duality,''
Adv.\ Theor.\ Math.\ Phys.\  {\bf 5}, 213 (2002)
[arXiv:hep-th/0108229].
%%CITATION = 00203,5,213;%%
}

%\MorrisonBM
\lref\MorrisonBM{
D.~R.~Morrison and J.~Walcher,
``D-branes and Normal Functions,''
arXiv:0709.4028 [hep-th].
%%CITATION = ARXIV:0709.4028;%%
}

%\OoguriBV
\lref\OoguriBV{
H.~Ooguri and C.~Vafa,
``Knot invariants and topological strings,''
Nucl.\ Phys.\  B {\bf 577}, 419 (2000)
[arXiv:hep-th/9912123].
%%CITATION = NUPHA,B577,419;%%
}

%\PolchinskiMT
\lref\PolchinskiMT{
J.~Polchinski,
``Dirichlet-Branes and Ramond-Ramond Charges,''
Phys.\ Rev.\ Lett.\  {\bf 75}, 4724 (1995)
[arXiv:hep-th/9510017].
%%CITATION = PRLTA,75,4724;%%
}

%RainvilleBook
\lref\RainvilleBook{
E.~Rainville,
``Special Functions,"
The Macmillan Company, New York 1960.
}

%\StromingerPD
\lref\StromingerPD{
A.~Strominger,
``Special Geometry,''
Commun.\ Math.\ Phys.\  {\bf 133}, 163 (1990).
%%CITATION = CMPHA,133,163;%%
}

%\TaylorII
\lref\TaylorII{
T.~R.~Taylor and C.~Vafa,
``RR flux on Calabi-Yau and partial supersymmetry breaking,''
Phys.\ Lett.\  B {\bf 474}, 130 (2000)
[arXiv:hep-th/9912152].
%%CITATION = PHLTA,B474,130;%%
}

%\VoisinAA
\lref\VoisinAA{
C.~Voisin,
``Hodge Theory and Complex Algebraic Geometry II,''
Cambridge University Press, Cambridge 2003.
}

%\WalcherRS
\lref\WalcherRS{
J.~Walcher,
``Opening mirror symmetry on the quintic,''
Commun.\ Math.\ Phys.\  {\bf 276}, 671 (2007)
[arXiv:hep-th/0605162].
%%CITATION = CMPHA,276,671;%%
}

%\WittenEP
\lref\WittenEP{
E.~Witten,
``Branes and the dynamics of {QCD},''
Nucl.\ Phys.\  B {\bf 507}, 658 (1997)
[arXiv:hep-th/9706109].
%%CITATION = NUPHA,B507,658;%%
}

%\WittenFB
\lref\WittenFB{
E.~Witten,
``Chern-Simons Gauge Theory As A String Theory,''
Prog.\ Math.\  {\bf 133}, 637 (1995)
[arXiv:hep-th/9207094].
%%CITATION = PMTMA,133,637;%%
}

%\WittenYC
\lref\WittenYC{
E.~Witten,
``Phases of N = 2 theories in two dimensions,''
Nucl.\ Phys.\  B {\bf 403}, 159 (1993)
[arXiv:hep-th/9301042].
%%CITATION = NUPHA,B403,159;%%
}

%\WittenZZ
\lref\WittenZZ{
E.~Witten,
``Mirror manifolds and topological field theory,''
arXiv:hep-th/9112056.
%%CITATION = HEP-TH/9112056;%%
}

%%%%%%%%%%%%%%%%%%%%%%
%  end references
%%%%%%%%%%%%%%%%%%%%%%

%%%%%%%%%%%%%%%   Title Page  %%%%%%%%%%%%%
\Title{\vbox{\hbox{\tt SU-ITP-08/14}} }
{\vbox{
  \centerline{\hbox{Effective superpotentials}}
  \vskip 0.7cm
  \centerline{\hbox{for compact D5-brane Calabi-Yau geometries}}
}}
\bigskip
\centerline{\bf Hans Jockers\footnote{${}^{\,a}$}{\tt jockers@stanford.edu}
                       and Masoud Soroush\footnote{${}^{\,b}$}{\tt soroush@stanford.edu}}
\bigskip
\centerline{{\it Department of Physics, Stanford University}}
\centerline{{\it  Stanford, CA 94305-4060, USA}}
\bigskip

\vskip 0.5cm
\centerline{\bf Abstract}
For compact Calabi-Yau geometries with D5-branes we study $N=1$ effective superpotentials depending on both open- and closed-string fields. We develop methods to derive the open/closed Picard-Fuchs differential equations, which control D5-brane deformations as well as complex structure deformations of the compact Calabi-Yau space. Their solutions encode the flat open/closed coordinates and the effective superpotential. For two explicit examples of compact D5-brane Calabi-Yau hypersurface geometries we apply our techniques and express the calculated superpotentials in terms of flat open/closed coordinates. By evaluating these superpotentials at their critical points we reproduce the domain wall tensions that have recently appeared in the literature. Finally we extract orbifold disk invariants from the superpotentials, which, up to overall numerical normalizations, correspond to orbifold disk Gromov-Witten invariants in the mirror geometry.  

\bigskip
\Date{\sl {August, 2008}}
\vfill\eject

%%%%%%%%%%%%%%%%%%%%%%%%%%%%%%%%%%
\newsec{Introduction}
%%%%%%%%%%%%%%%%%%%%%%%%%%%%%%%%%%
Since D-branes have been discovered in string theory as non-perturbative BPS~objects~\PolchinskiMT, they have played an important role. Besides serving as crucial ingredients in phenomenological string model building they have increased our insight into non-perturbative physics in both string and field theory. Over and above D-branes have also deepened our understanding of the web of string dualities. Such dualities often map theories in the quantum regime to dual descriptions in which semiclassical methods are applicable. One prominent example of this kind is given by mirror symmetry, which connects classical geometry to a notion of quantum geometry. In the context of D-branes mirror symmetry is further refined and leads towards the homological mirror symmetry conjecture \refs{\KontsevichAA,\HoriAA}.

Mirror symmetry in the closed string sector relates type~IIA string theory compactified on a Calabi-Yau threefold, $X$, to type~IIB string theory compactified on the mirror Calabi-Yau threefold, $Y$.  Among other things the duality implies that the associated four-dimensional low energy effective $N=2$ supergravity theories are the same. In particular the target space manifolds of the scalar fields in the $N=2$ vector multiplets are captured by the same holomorphic prepotential. On the type~IIB side this holomorphic function is derived by analyzing the complex structure moduli space of the Calabi-Yau, $Y$, by means of classical geometric, whereas on the mirror type~IIA side the prepotential arises from the quantum geometry of the K\"ahler moduli space of the Calabi-Yau, $X$.

The holomorphic prepotential arises from the underlying $N=2$ special geometry, which is a consequence of the $N=2$ local supersymmetry of the compactified type~II string theories. Another approach to this $N=2$ special geometry structure appears by investigating the topological A- and B-model \WittenZZ. These topological string theories can be viewed as subsectors of the physical type~IIA and type~IIB string theories respectively \refs{\WittenZZ,\AntoniadisZE,\BershadskyCX}. Then mirror symmetry connects the A-model on the Calabi-Yau manifold, $X$, to the B-model on the Calabi-Yau manifold, $Y$. In this context mirror symmetry can be extended to the open-string sector by including topological branes. The homological mirror symmetry conjecture states that the category of topological B-branes in the B-model is equivalent to the category of topological A-branes in the mirror A-model \refs{\KontsevichAA,\DouglasGI,\LazaroiuJM,\AspinwallPU,\HoriAA}. Excellent reviews of these matters may be found in refs.~\refs{\LazaroiuMD,\AspinwallJR}.

Analogously to the closed-string sector one would like to take advantage of this extended version of mirror symmetry for explicit computational purposes in the open-string sector. As type~II string theories with branes compactified on Calabi-Yau manifolds exhibit only $N=1$ local supersymmetry their low-energy regime is given by four-dimensional $N=1$ supergravity theories. Part of the defining data of these supergravity theories is the holomorphic superpotential, on which we focus in this work.

Similarly as the prepotential in a purely closed-string setup it turns out that the $N=1$ superpotential of string compactifications with branes and fluxes are given on the type~IIB side by classical obstruction theory, whereas on the type~IIA side they are generated non-perturbatively by open-string disk instantons \refs{\WittenFB,\KachruIH}. Thus from the $N=1$ superpotential of the topological B-model we get a handle on the non-perturbative superpotential  of the topological A-model for the mirror configuration.

As the superpotential corresponds in the topological A-model to the disk partition function \refs{\OoguriBV,\LabastidaZP}, open mirror symmetry provides a powerful tool in enumerative geometry. In the large radius region of the topological A-model the disk partition function counts integer disk invariants \OoguriBV, whereas, as investigated recently in refs.~\refs{\BouchardYS,\BriniRH,\BouchardGU}, in the vicinity of orbifold points the superpotential encodes rational orbifold disk invariants \refs{\CoatesAA,\BayerAA,\BouchardAA}. 

Moreover, the effective superpotential is a relevant quantity not only in enumerative geometry but also in string phenomenology. It constitutes an important ingredient in string model building because, for instance, it stabilizes moduli fields and/or triggers supersymmetry breaking. Most of the phenomenological $N=1$ type~II string models are either constructed from non-compact geometries, and hence gravity is decoupled, or they are obtained by semi-classical Kaluza-Klein reductions, for which the explicit quantum corrections are often not known. In order to capture some of these corrections in the context of compact Calabi-Yau scenarios it is desirable to get at least a handle on the quantum-corrected $N=1$ effective superpotential.

In practice computing such effective superpotential is rather hard. First of all for a given brane configuration in a compact Calabi-Yau manifold on the B-model side the corresponding brane configuration on the mirror A-model side is only known in special situations. Second even if the mirror configurations are known on both sides one still needs to find the open-closed string mirror map, which provides for the dictionary translating the classical computation in the topological B-model to the quantum computation in the topological A-model.

For non-compact Calabi-Yau manifolds the program sketched above has successfully been carried out in refs.~\refs{\AganagicGS,\AganagicNX,\LercheCK,\LercheYW,\AganagicDB}. In ref.~\refs{\AganagicGS,\AganagicNX} the boundary condition at infinity has been used, whereas in refs.~\refs{\LercheCK,\LercheYW} the concept of $N=1$ special geometry has been introduced and applied. Although the connection between these two approaches is not obvious both methods yield results which are in agreement.

For compact Calabi-Yau manifolds the analog analysis seems to be more involved and to our knowledge the above sketched endeavor, namely to compute quantum corrected superpotentials depending on both open- and closed-string moduli has not been carried out explicitly so far. Recently, however, in refs.~\refs{\WalcherRS,\MorrisonBM} a major step in this direction has been achieved by computing quantum-corrected domain wall tensions on the quintic threefold using open-string mirror symmetry. Following this recipe similar computations have been carried out successfully for other one-parameter Calabi-Yau geometries in refs.~\refs{\KnappUW,\KreflSJ}.

Guided by the $N=1$ special geometry techniques applied to non-compact Calabi-Yau geometries in refs.~\refs{\LercheCK,\LercheYW}, we propose in this work an analog method to derive Picard-Fuchs equations governing effective superpotentials for D5-brane configurations in compact Calabi-Yau manifolds. The resulting Picard-Fuchs partial differential equations depend on both open- and closed-string moduli, and their solutions encode in addition to the effective superpotential the open-/closed-string mirror map. Thus our approach is not only suitable to compute effective superpotentials but also to extract enumerative invariants in the topological A-model of the mirror geometry.

We apply these novel techniques to D5-branes in Calabi-Yau threefolds, which are related to the geometries discussed in the context of domain wall tensions in refs.~\refs{\WalcherRS,\KnappUW,\KreflSJ}. For two examples we explicitly derive the quantum corrected superpotential, which we express in the vicinity of the orbifold point by a uniquely distinguished set of open/closed flat coordinates. Then, similarly to the analysis performed for D-branes in local Calabi-Yau geometries \refs{\BouchardYS,\BriniRH,\BouchardGU}, we extract (up to overall numerical normalizations) from the flat superpotential a tower of orbifold disk invariants for the mirror D-brane configuration in the compact mirror Calabi-Yau geometry. Finally, as a bonus and as a highly non-trivial consistency check we reproduce for our two examples the domain wall tension computed in refs.~\refs{\WalcherRS,\KnappUW,\KreflSJ} by evaluating the calculated effective superpotential at its critical points.

The outline of this paper is as follows. In Section~2 we review some relevant aspects of $N=1$ special geometry along the lines of refs.~\refs{\LercheCK,\LercheYW}. Then in Section~3 we develop the tools to derive the open/closed Picard-Fuchs differential equations, and we argue that their solutions capture the necessary information to extract the effective superpotential in terms of flat coordinates. In Section~4 and Section~5 we apply our techniques in detail to two concrete examples. The first example is given by a family of D5-branes in the degree eight Calabi-Yau hypersurface of the weighted projective space, $\IW\IP^4_{(1,1,1,1,4)}/(\IZ_8)^2\times\IZ_2$, whereas the second example is a family of D5-branes in the mirror quintic Calabi-Yau threefold. For these examples we extract orbifold disk invariants for the associated mirror geometries and determine domain wall tensions in agreement with the results in the literature. Finally we present our conclusions in Section~5. In the two appendices we supplement further computational details for the two discussed examples.

%%%%%%%%%%%%%%%%%%%%%%
\newsec{Effective superpotentials and $N=1$ special geometry}
%%%%%%%%%%%%%%%%%%%%%%
It is well-known that type~II string theory compactified on Calabi-Yau threefolds with background fluxes or space-time filling D-branes is described in the low-energy regime by $N=1$ effective supergravity theories. For such string compactifications we present in this work new techniques to compute the effective superpotential, which is part of the defining data of $N=1$ supergravities and plays an important role in many phenomenological applications. To set the stage for the subsequent sections we first review some aspects of $N=1$ special geometry, which are relevant for our computations.

%%%%%%%%%%%%%%%%%%%%%%
\subsec{Flux-induced and D5-brane superpotentials}
\subseclab\secFluxBraneSuper
%%%%%%%%%%%%%%%%%%%%%%
Let us consider type~IIB string theory compactified on the Calabi-Yau threefold, $Y$. Then in the presence of internal background fluxes an effective superpotential is induced. Here we mainly focus on the quantized three-from RR~fluxes, $F^{(3)}$, which takes values in the integer cohomology group, $H^3(Y,\IZ)$. Then the resulting superpotential reads \refs{\TaylorII,\GukovYA}
\eqn\RRW{ W_{\rm RR}(z) \,=\, \int_Y  \Omega(z) \wedge F^{(3)} \ , }
where $\Omega(z)$ is up to normalization the unique holomorphic three form of the Calabi-Yau threefold, $Y$, depending on the complex structure moduli parametrized by the coordinates, $z$.\foot{Strictly speaking a modulus parametrizes a flat direction in the scalar potential of the effective field theory. In this paper, however, a modulus refers to the complex scalar field of a neutral chiral multiplet, which may or may not be obstructed by the superpotential.} The dependence on the complex structure moduli can be made more explicit by expressing the three-form superpotential in terms of the period vector, $\Pi^\alpha(z)$, of the Calabi-Yau manifold, which is obtained by integrating the holomorphic three form, $\Omega$, over a basis, $\Gamma_\alpha$, of the integer homology group, $H_3(Y,\IZ)$,
\eqn\BulkPeriods{\Pi^\alpha(z)\,=\,\int_{\Gamma_\alpha} \Omega(z) \ ,  \quad \Gamma_\alpha \in H_3(Y,\IZ) \ . }
The periods, $\Pi^\alpha(z)$, of a Calabi-Yau manifold are governed by the underlying $N=2$ special geometry, which gives rise to the holomorphic prepotential of the vector multiplets in the associated $N=2$ supergravity theory. Here we also express the flux-induced superpotential, $W_{\rm RR}$, in terms of these periods 
\eqn\RRW{W_{\rm RR}(z) \,=\, N_\alpha\, \Pi^\alpha(z) \ , }
where the integers, $N_\alpha$, are the quanta of the three-form background fluxes, $F^{(3)}$.

A similar effective superpotential arises from space-time filling D-branes wrapping even-dimensional cycles in type~IIB Calabi-Yau compactifications. For branes filling the whole compactification space the open-string partition function, which in our context yields the resulting effective superpotential, arises from the holomorphic Chern-Simons action \WittenFB. As we focus in this work on a space-time filling D5-brane wrapping a two cycle, $C$, of the internal Calabi-Yau, $Y$, we need to consider the dimensional reduction of the holomorphic Chern-Simons action to two dimensions, which becomes \refs{\WittenEP,\KachruIH,\AganagicGS}
\eqn\DFiveW{
  W_{\rm D5} \,=\, \int_C \Omega_{ijw}\, \zeta^i \bar\partial_w \zeta^j \, dw\,d\bar w \ . }
Here $\zeta^i$ are sections of the normal bundle of the two cycle, $C$, embedded in its ambient Calabi-Yau space, $Y$, and they parametrize infinitesimal deformations of the D5-branes. The holomorphic Chern-Simons action also depends on the complex structure moduli through its coupling to the holomorphic three form, $\Omega$, and therefore the superpotential, $W_{\rm D5}$, depends on both the complex structure moduli and the D5-brane open-string moduli for the deformations of the embedding cycle, $C$.

Analogously to the flux-induced superpotential the moduli dependence becomes explicit by writing the D5-brane superpotential in terms of semi periods. The semi period vector, $\hat\Pi^{\hat\alpha}$, is defined by
\eqn\DFivePeriods{\hat\Pi^{\hat\alpha}(z,u)\,=\,\int_{\hat\Gamma^{\hat\alpha}(u)} \Omega(z) \ , }
where $\hat\Gamma^{\hat\alpha}(u)$ constitutes a basis of three chains that have non-trivial boundaries, $\partial\hat\Gamma^{\hat\alpha}(u)$, lying in the union of non-trivial two-cycles, $S$, of the Calabi-Yau manifold, $Y$. As this basis of open three chains depends on the open-string moduli, $u$, the semi period vector, $\hat\Pi^{\hat\alpha}(z,u)$, becomes a function of both closed and open fields, and the moduli-dependent D5-brane superpotential~\DFiveW\  reads \refs{\WittenEP,\AganagicGS,\MayrXK}
\eqn\DFiveWPeriod{W_{\rm D5}(z,u)\,=\, \hat N_{\hat\alpha}\, \hat\Pi^{\hat\alpha}(z,u) \ . }
Here the integers, $\hat N_{\hat\alpha}$, specify the topology of the internal two cycle, $C$, of the D5-brane worldvolume by specifying a linear combination of the the two cycles in the set, $S$.

Since both the flux-induced and the D5-brane superpotential arise from integrals of the holomorphic three form, $\Omega$, it is natural to consider the combined superpotential \refs{\LercheCK,\LercheYW}
\eqn\WRelPeriod{W(z,u)\,=\, W_{\rm RR}(z)+W_{\rm D5}(z,u) \,=\, \rel{N}_a\,\rel{\Pi}^a(z,u) \ , }
in terms of the relative period vector
\eqn\RelPeriods{\rel{\Pi}^a(z,u) \,=\, \int_{\rel{\Gamma}^a(u)} \Omega(z) \ , \quad
  \rel{\Gamma}^a(u) \in H_3(Y,S,\IZ) \ . }
Here $\rel{\Gamma}^a$ denotes a basis of three chains in the relative integer homology group, $H_3(Y,S,\IZ)$. We should stress that the basis, $\rel{\Gamma}^a$, captures closed three chains, $\Gamma^\alpha$, and open three chains, $\hat\Gamma^{\hat\alpha}$, with boundaries in the set of two cycles, $S$. Therefore the effective superpotential~\WRelPeriod\ does indeed get contributions from three-form RR~fluxes and D5-branes, and the integers, $\rel{N}_a$, specify now both the three-form flux quanta and the D5-brane topology. Note that also from a physics point of view the interplay of three-form fluxes and space-time filling D5-branes in the effective superpotential is not very surprising because D5-branes and three-form fluxes are often related by geometric transitions \CachazoJY. 

Although the effective superpotential~\WRelPeriod\ is a purely classical expression for type~IIB string compactifications it describes a highly intricate sum of non-perturbative instantons for the mirror type~IIA string compactification \KachruAN. In order to extract these non-trivial instanton contributions it is necessary to analyze the structure of these relative periods. Analogously to the $N=2$ special geometry, which relates the holomorphic $N=2$ prepotential to periods of Calabi-Yau manifolds, the relative periods entering the $N=1$ holomorphic superpotential~\WRelPeriod\  are governed by the underlying $N=1$ special geometry, which has been introduced in refs.~\refs{\MayrXK,\LercheCK,\LercheYW}.

However, before discussing the properties of relative periods there are a few general comments in order. First of all since we are interested in effective superpotentials arising from compact Calabi-Yau geometries the background three-form fluxes and the space-time filling D5-branes introduce RR~tadpoles rendering in the physical string theory inconsistent. These tadpoles arise from worldsheets at the one-loop level and can be cancelled by introducing appropriate orientifold planes. But these orientifold planes do not alter our computations because the effective superpotential~\WRelPeriod\ appears at string tree level. Moreover, as a BPS protected quantity the superpotential is also not further modified by flux- or brane-induced backreactions to the geometry. Finally we remark that due to the $SL(2,\IZ)$ symmetry of type~IIB string theory the effective superpotential~\WRelPeriod\ can easily be extended to describe NS three-form fluxes and NS5-branes \refs{\LercheCK,\LercheYW}. This is achieved by replacing the RR~sector charges, $\rel{N}_a$, by the complexified charge quanta, $\rel{N}_a+\tau \rel{N}_a^{\rm NS}$. Here $\tau$ denotes the complex dilaton and the integers, $\rel{N}_a^{\rm NS}$, capture the NS~sector charges. However, in the following we set the NS~charges again to zero and restrict ourselves to the RR~sector.

%%%%%%%%%%%%%%%%%%%%%%
\subsec{Relative periods}
%%%%%%%%%%%%%%%%%%%%%%
As relative periods are crucial for the effective superpotentials of interest we study now their structure in some more detail. These relative periods are adequately described in terms of relative homology and relative cohomology \refs{\LercheCK,\LercheYW}. Therefore we give a brief mathematical interlude to these matters. For more details see, for instance, ref.~\KaroubiBook.

For the submanifold, $S$, embedded by the map, $i: S \hookrightarrow Y$, in the ambient Calabi-Yau manifold, $Y$, the space of relative forms, $\Omega^*(Y,S)$, is the subspace of forms, $\Omega^*(Y)$, defined as the kernel of the pullback, $i^*: \Omega^*(Y)\rightarrow \Omega^*(S)$. In other words the relative forms fit in the exact sequence
\eqn\ShortSeq{
  0  \longrightarrow \Omega^*(Y,S) \hookrightarrow 
  \Omega^*(Y) {\buildrel i^* \over \longrightarrow} \Omega^*(S)
  \longrightarrow 0 \ . }
Then the relative cohomology groups, $H^*(Y,S)$, arise from the space of closed modulo exact relative forms with respect to the de~Rham differential, $d$. Since the de~Rham differential commutes with the maps in the short exact sequence~\ShortSeq, we deduce a long exact sequence on the level of cohomology in the usual manner. In particular the long exact sequence implies for the three-form cohomology group, $H^3(Y,S)$,  
\eqn\HThreeRel{
  H^3(Y,S) \, \cong \, {\rm ker}\left(H^3(Y)\rightarrow H^3(S)\right)\oplus
  {\rm coker}\left(H^2(Y)\rightarrow H^2(S)\right) \ . }
For Calabi-Yau threefolds the first summand equals $H^3(Y)$ on dimensional grounds. The second contribution is the variable cohomology group, $H^2_{\rm var}(S)$, with respect to the embedding space, $Y$, \ie these are cohomology elements of the submanifold not induced form the ambient space. Consequently the decomposition~\HThreeRel\ allows us to represent a relative three form, $\rel{\Xi}$, as a pair of a closed three form, $\Xi$, and a closed two form, $\xi$, 
\eqn\RelThree{\rel{\Xi} \,=\, \left(\Xi, \xi \right) \,\in\, H^3(Y,S) \ ,}
such that the relative form, $\rel{\Xi}$, obeys the equivalence relation
\eqn\RelThreeEquiv{\rel{\Xi} \, \sim \, \rel{\Xi} + \left(d\alpha, i^*\alpha-d\beta \right) \ . }
Here $\alpha$ is a two form on the Calabi-Yau, $Y$, whereas $\beta$ is a one form on the subspace, $S$.

Let us turn to the homology group, $H_3(Y,S,\IZ)$, of relative three cycles. As mentioned previously a relative three cycle, $\rel{\Gamma}$, is a three chain whose boundary, $\partial\rel{\Gamma}$, lies in the submanifold, $S$. The duality pairing between relative three-form cohomology elements and relative three cycles is given by
\eqn\RelPairing{\int_{\rel{\Gamma}} \rel{\Xi} \,\equiv\,
   \int_{\rel{\Gamma}} \Xi - \int_{\partial\rel{\Gamma}} \xi \ . }
This topological pairing is compatible with the equivalence relation~\RelThreeEquiv, and hence it is well-defined. 

After this brief introduction to relative (co-)homology we can write the relative periods~\RelPeriods\ in terms of the relative three form, $\rel{\Omega}(z,u)$, integrated over a relative homology basis, $\rel{\Gamma}^a$,
\eqn\RelPeriodsTwo{ \rel{\Pi}^a(z,u)\,=\, \int_{\rel{\Gamma}^a} \rel{\Omega}(z,u) \ , \quad
  \rel{\Gamma}^a \in H_3(Y,S,\IZ) \ . }
At a given reference point in the open-/closed-string moduli space the relative three form, $\rel{\Omega}$, can be viewed as the pair, $(\Omega,0)$. However, as we move in the moduli space the relative three form changes and generically acquires also non-zero two-form contributions. The moduli dependence of the relative periods, $\rel{\Pi}^a(z,u)$, is now entirely captured by the variation of the relative form, $\rel{\Omega}(z,u)$.

%%%%%%%%%%%%%%%%%%%%%%
\subsec{Variation of mixed Hodge structure}
\subseclab\secMixedHodge
%%%%%%%%%%%%%%%%%%%%%%
The suitable formalism to get a handle on the moduli dependence of the relative periods~\RelPeriodsTwo\ is the variation of mixed Hodge structure \refs{\LercheCK,\LercheYW}. It is a generalization of the variation of Hodge structure used to compute the complex structure moduli dependence of the closed-string period vector. 

For the mathematical definition of a mixed Hodge structure we refer the reader to refs.~\refs{\DeligneII,\VoisinAA}. In our context, relative three forms realize a mixed Hodge structure as follows \refs{\LercheCK,\LercheYW}: The $\IZ$-module of a mixed Hodge structure is given by the relative integer cohomology group, $H^3(Y,S,\IZ)$. The second ingredient is a finite decreasing filtration, $F^p$, of the complexified group, $H^3(Y,S,\IC) \equiv \IC \otimes_{\IZ} H^3(Y,S,\IZ)$. This filtration becomes
\eqn\FFilt{\eqalign{
  F^3 \,&=\, H^{3,0}(Y,S) \ , \cr 
  F^2 \,&=\, H^{3,0}(Y,S) \oplus H^{2,1}(Y,S) \ , \cr
  F^1 \,&=\, H^{3,0}(Y,S) \oplus H^{2,1}(Y,S) \oplus H^{1,2}(Y,S) \ , \cr 
  F^0 \,&=\, H^{3,0}(Y,S) \oplus H^{2,1}(Y,S) \oplus H^{1,2}(Y,S)\oplus H^{0,3}(Y,S) \ . }}
In terms of the decomposition~\HThreeRel\ the groups, $H^{p,q}(Y,S)$, split into the cohomology groups, $H^{p,q}(Y)$ and $H^{p,q-1}_{\rm var}(S)$. Thus in particular the holomorphic three form, $\Omega$, spans the filtration, $F^3$. Finally a mixed Hodge structure has a finite increasing weight filtration, $W_p$, on the rational relative cohomology group, $H^3(Y,S,\IQ) \equiv  \IQ \otimes_{\IZ} H^3(Y,S,\IZ)$. The weight filtration is again induced from the decomposition~\HThreeRel, and it reads
\eqn\WFilt{
  W_3 \,\cong\, H^3(Y,\IQ) \ , \quad 
  W_4 \,\cong\, H^3(Y,\IQ)\oplus H^2_{\rm var}(S,\IQ)\,\cong\, H^3(Y,S,\IQ) \ . }
Note that the finite decreasing filtration, $\tilde F^p\equiv W_3\cap F^p$, gives rise to the Hodge structure, 
\eqn\WFiltBulk{\tilde F^p\,=\,\bigoplus_{k=0}^{3-p} H^{3-k,k}(Y) \ , \quad p=0,1,2,3 \ , }
associated to the closed-string complex structure moduli space.

In order to analyze the moduli-dependent relative period vector~\RelPeriodsTwo, we discuss the behavior of relative three forms under infinitesimal deformations. It is well-known that an infinitesimal closed-string complex structure deformation, $\partial_z$, changes the Hodge type of a $(p,q)$-form. On the other hand an infinitesimal open-string deformation, $\partial_u$, does not modify the closed-string periods because it gives rise to an infinitesimal deformation of the normal bundle of the submanifold, $S$, which only affects the two-form sector, $H^2_{\rm var}(Y)$. Thus, as has been shown rigorously in ref.~\LercheYW , the infinitesimal deformations, $\partial_z$ and $\partial_u$, viewed as tangent vectors in the open-/closed string moduli space, schematically act on the defined mixed Hodge structure as:
\eqn\VarMixed{
\xymatrix{
  F^3\cap W_3 \ar[r]^{\partial_z} \ar[rd]^{\partial_u}& F^2\cap W_3  \ar[r]^{\partial_z}  \ar[rd]^{\partial_u}
     & F^1\cap W_3  \ar[r]^{\partial_z} \ar[rd]^{\partial_u} & F^0\cap W_3 \ar[d]^{\partial_u} \cr  
  & F^2\cap W_4 \ar[r]^{\partial_z,\partial_u} &  F^1\cap W_4 \ar[r]^{\partial_z,\partial_u} & F^0 \cap W_4 } \ . }
Note that the two-form sector, $H^2_{\rm var}(Y) \cong W_4 / W_3$, constitutes a sub-system, 
\eqn\VarSubSys{ 
   F^2\cap (W_4/W_3)\, {\buildrel\partial_z,\partial_u\over\longrightarrow} \,
   F^1\cap (W_4/W_3)\,{\buildrel\partial_z,\partial_u\over\longrightarrow} \, F^0 \cap (W_4/W_3) \ , }
which is closed with respect to the variations, $\partial_z$ and $\partial_u$, and which will play an important role in deriving and solving the Picard-Fuchs differential equations of the relative periods~\RelPeriodsTwo. 

The variation of mixed Hodge structure exhibits the $N=1$ special geometry of the open-/closed-string moduli space. As has been pointed out in ref.~\refs{\LercheCK,\LercheYW}, we should view the emerging structures as a distinguished feature of $N=1$ supergravity theories arising from $N=1$ string compactifications and not as a property of a generic $N=1$ supergravity theory.

%%%%%%%%%%%%%%%%%%%%%%
\newsec{Picard-Fuchs equations for D5-branes in compact Calabi-Yau geometries}
%%%%%%%%%%%%%%%%%%%%%%
In this section we develop the machinery to compute effective superpotentials of D5-branes in compact Calabi-Yau threefolds. We focus on D5-brane geometries whose moduli spaces are describable by studying certain divisors of the embedding Calabi-Yau spaces. Furthermore these Calabi-Yau threefolds are realized as hypersurfaces in four-dimensional complex (weighted) projective spaces. The main idea is to express the relative three forms associated to the D5-brane geometry in terms of residue integrals. Since both open- and closed-string moduli enter in the definition of these residue integrals, we get a direct handle on the moduli dependence of relative three forms. Hence we are able to explicitly analyze the corresponding variation of mixed Hodge structure, which then allows us to derive the Picard-Fuchs equations of the relative period vector governing the effective superpotential.

%%%%%%%%%%%%%%%%%%%%%%
\subsec{Residue integrals for three forms in Calabi-Yau threefolds}
\subseclab\secResThree
%%%%%%%%%%%%%%%%%%%%%%
Before we construct the mixed Hodge filtration for relative forms we first recall how to describe three forms of a Calabi-Yau threefold by means of residue integrals \refs{\GriffithsAA,\CandelasSE,\LercheWM}. In the following the Calabi-Yau hypersurface, $Y$, is given as the zero locus, $P\equiv 0$, of a (quasi-) homogeneous polynomial, $P$, in the complex (weighted) projective space $\IW\IP^4_{(a_1,a_2,a_3,a_4,a_5)}$. In order for the hypersurface, $Y$, to be Calabi-Yau the defining polynomial, $P$, must be (quasi-)homogeneous of degree $d=a_1+\ldots+a_5$.

By integrating over a tubular neighborhood, $\gamma$, of the zero locus of the polynomial, $P$, the residue integral, 
\eqn\ThreeForms{\Xi^k\,=\,\int_\gamma{p(x) \over P^{k+1}} \Delta \ , }
yields a three form, $\Xi^k$, on the Calabi-Yau threefold, $Y$. Here $p(x)$ is a (quasi-)homogeneous polynomial of degree $d\,k$ of the projective coordinates $[x_1 : x_2 : x_3 : x_4 : x_5]$, whereas the differential, $\Delta$, is given by \GriffithsAA
\eqn\WPDelta{ \Delta\,=\,\sum_{m=1}^5 (-1)^m\, a_m\,x_m\,
  dx_1\wedge \ldots \widehat{dx_m} \ldots \wedge dx_5  \ . }
In this sum the differentials, $\widehat{dx_m}$, are omitted as indicated by the hat, and the residue integral~\ThreeForms\ is well-defined as it is invariant under quasi-homogeneous rescaling. Note that by acting with the de~Rham differential, $d$, it is easy to see that the three form, $\Xi^k$, is closed.

Next we turn our attention to exact forms. First we observe that we can represent two forms with residue integrals
\eqn\TwoForms{
  \alpha\,=\,\int_\gamma \sum_{m<n} {a_n\,x_n\,q_m(x) - a_m\,x_m\,q_n(x)\over P^k}
  (-1)^{m+n} dx_1\wedge\ldots\widehat{dx_m}\ldots\widehat{dx_n}\ldots\wedge dx_5 \ . }
The polynomials, $q_n(x)$, have degree $d(k-1)+a_n$. Acting with the differential, $d$, on the residue integral~\TwoForms\ we obtain the exact three form
\eqn\dalpha{
  d\alpha\,=\,\int_\gamma \sum_m \left[k{q_m \partial_m P\over P^{k+1}}
  -{\partial_m q_m\over P^k} \right] \Delta \ . }

Thus we have assembled all the ingredients to represent the de~Rham three-form cohomology elements. As shown in ref.~\GriffithsAA\ and as suggested by the structure of the exact forms~\dalpha\ each non-trivial element of the polynomial ring, $\IC[x]/(\partial_n P)$, of degree $d\,k$ corresponds to a distinct non-trivial element, $\Xi^k$, in the cohomology group, $H^3(Y,\IC)$.\foot{For weighted projective spaces the residue integrals~\ThreeForms\ do not always span the whole cohomology group, $H^3(Y,\IC)$. With residue integrals we can only describe those three-form cohomology elements that correspond to toric divisors in the mirror geometry.}  

A refined analysis reveals that a three form, $\Xi^k$, of grade $k$, arising from a polynomial of degree $d\,k$, lies in the Hodge filtration module, $\tilde F^{3-k}$, defined in eq.~\WFiltBulk\ \GriffithsAA. In particular the unique holomorphic three form, $\Omega$, of the Calabi-Yau hypersurface, $Y$, which spans the filtration, $\tilde F^3$, is given by
\eqn\HolOmega{\Omega\,=\,\int_\gamma {1 \over P} \Delta \ . }
This expression allows us to investigate the complex structure dependence of the holomorphic three form, $\Omega$, by considering a family of hypersurface polynomials, $P(z)$, parametrized by the moduli, $z$. Moreover by taking $k$-th order derivatives with respect to the parameters, $z$, we realize infinitesimal deformations of order~$k$ and obtain three forms at grade~$k$. Thus we are able to explicitly study the variation of Hodge structure,
\eqn\VarBulk{ 
  \tilde F^3\,{\buildrel \partial_z\over\longrightarrow}\,
  \tilde F^2\,{\buildrel \partial_z\over\longrightarrow}\,
  \tilde F^1\,{\buildrel \partial_z\over\longrightarrow}\,\tilde F^0 \ , }
of the closed-string complex structure moduli space.  

So far we have argued that a three form, $\Xi^k$, lies in the filtration module, $\tilde F^{3-k}$. If, however, the associated polynomial, $p(x)$, of degree $d\,k$, lies in the ideal of the polynomial ring, $\IC[x]/(\partial_n P)$, \ie $p(x) \equiv \partial_m q_m(x)$ for some polynomial, $q_m(x)$, then according to eq.~\dalpha\ we can add an appropriate exact form, $d\alpha$, such that the three form is reduced to a three form at grade~$k-1$. Hence we have shown that the three form, $\Xi^k$, is even an element of the filtration module, $\tilde F^{4-k}$. Recursively repeating this process eventually we either arrive at some non-trivial three-form cohomology element at lower grade or the final defining polynomial becomes zero. In the latter situation we have establish that the original three form, $\Xi^k$, is exact and thus trivial in cohomology. This reduction method is sometimes called the Griffiths-Dwork algorithm. Later we will use a generalization of this algorithm to derive the Picard-Fuchs equations for the relative periods.

%%%%%%%%%%%%%%%%%%%%%%
\subsec{Residue integrals for relative three forms in Calabi-Yau threefolds}
\subseclab\secResRelThree
%%%%%%%%%%%%%%%%%%%%%%
In order to apply the concepts of $N=1$ special geometry to D5-branes in compact Calabi-Yau threefolds the next task is to establish residue integral representations for relative three forms. These integrals are derived by exploiting the relative three-form decomposition~\HThreeRel\ into a two-form/three-form pair. After having thoroughly explored the three form piece in Section~\secResThree\ we first focus now on the variable two-form cohomology, $H^2_{\rm var}(S)$, which captures the open-string moduli dependence of the embedded D5-brane. 

Since we want to study D5-brane effective superpotentials the two cycle wrapped by the D5-brane is generically not holomorphic. In fact it is only holomorphic if the moduli coincide with a critical point of the superpotential. To avoid the complication of dealing with non-holomorphic submanifolds we employ the arguments of refs.~\refs{\LercheCK,\LercheYW} and replace the submanifold, $S$, by a holomorphic hypersurface, $V$, of the Calabi-Yau manifold, $Y$,  such that this four-dimensional space embeds the wrapped cycles. One might be worried that the replacement introduces additional structure not related to the D5-brane geometry. However, we will see that for the examples discussed in this work, this substitution process does not give rise to fake additional moduli for the computed relative periods. Therefore we assume in the following that it is possible to use the simpler cohomology group, $H^3(Y,V)$, instead of its complicated ancestor, $H^3(Y,S)$. In particular the two-form part of the relative three forms are now captured by the variable cohomology, $H^2_{\rm var}(V)$.

In practice the holomorphic four-dimensional subspace, $V$, is associated to a divisor of the Calabi-Yau space, $Y$, \ie the manifold, $V$, arises as the zero locus, $Q\equiv 0$, of the (quasi-)homegenous polynomial, $Q$, of degree~$f$. Hence we can view the subspace, $V$, as the complete intersection of the Calabi-Yau polynomial, $P$, and the D5-brane polynomial, $Q$, in the (weighted) projective space, $\IW\IP^4_{(a_1,a_2,a_3,a_4,a_5)}$. Therefore the residue integral, 
\eqn\TwoForms{\xi^{k+l}\,=\,\int_{\hat\gamma} {p(x) \over P^{k+1} Q^\ell}\Delta \ , }
represents a closed two form, $\xi^{k+l}$, of the manifold, $V$ \LibgoberAA. Here the differential, $\Delta$, is given by eq.~\WPDelta, and we integrate over the tubular neighborhood, $\hat\gamma$, of the intersection, $\{P\equiv 0\}\cap\{Q\equiv0\}$. The polynomial, $p(x)$, must have degree, $k\,d+\ell\,f$, so as to render the residue integral invariant under (quasi-)homogeneous rescaling as required for consistency reasons. The resulting form, $\xi^{k+l}$, can only be non-zero for $k\ge 0$ and $\ell>0$.

Analogously to eq.~\TwoForms\ the (quasi-)homogeneous polynomials, $q_m(x)$, of degree $(k-1)d+\ell\,f+a_m$ give rise to the residue integral,
\eqn\OneForms{
  \beta\,=\, \int_{\hat\gamma}
  \sum_{m<n} {a_n\,x_n\,q_m(x) - a_m\,x_m\,q_n(x)\over P^kQ^\ell}
  (-1)^{m+n} dx_1\wedge\ldots\widehat{dx_m}\ldots\widehat{dx_n}\ldots\wedge dx_5 \ , }
representing a one form on the submanifold, $V$. Acting with the de Rham differential, $d$, on the one form, $\beta$, we arrive after a few steps of algebra at the exact two form
\eqn\dbeta{d\beta\,=\,\int_{\hat\gamma}
\sum_m \left[k{q_m(x) \partial_m P\over P^{k+1}Q^\ell}
+\ell{q_m(x)\partial_m Q\over P^kQ^{\ell+1}}-{\partial_m q_m(x)\over P^k Q^\ell} \right] \Delta \ . }

The exact and the closed two forms enable us to study the cohomology resulting from the residue integrals. As shown in ref.~\GriffithsBB\ the residue integrals~\TwoForms\ capture non-trivial cohomology elements in the variable cohomology, $H^2_{\rm var}(V)$, which is precisely the cohomology group relevant in the decomposition~\HThreeRel\ of the relative cohomology group, $H^3(Y,V)$.\foot{This is a consequence of the Hard Lefschetz theorem.}  Moreover, the grades of the two forms~\TwoForms\ are compatible with the Hodge filtration of the variable two-form cohomology, \ie the two form, $\xi^k$, lies in the filtration module, $F^{3-k}\cap(W_4/W_3)$, of the mixed Hodge structure introduced in Section~\secMixedHodge. Altogether the two-form residue integrals provide for a suitable tool to investigate the variation of Hodge structure~\VarSubSys\ of the variable cohomology, $H^2_{\rm var}(V)$.

So far we have separately discussed the three- and two-form part in the decomposition~\WFilt\ of relative cohomology elements. However, in order to capture the variation of mixed Hodge structure~\VarMixed\ we need to take adequately into account the interplay of these two components. First we must incorporate the equivalence relation \RelThreeEquiv. Therefore we are required to derive for two forms the pullback, $i^*$, induced from the embedding, $i: V\hookrightarrow Y$, on the level of residue integrals. This is achieved by writing the pullback of the two form~\TwoForms\ in terms of an additional residue with respect to the zero locus of the divisor, $V$. A few steps of algebra reveal for the pullback two form~\TwoForms\foot{We have dropped an unimportant factor of $2\pi i$.}
\eqn\PBalpha{i^* \alpha\,=\,-\int_{\hat\gamma} \sum_m {q_m(x)\partial_m Q\over P^k Q}\Delta \ .}
Second we observe that if we introduce open-string moduli, $u$, by considering a family of divisors, $Q(u)$, this moduli dependence never enters the residue integral representation of the three forms~\ThreeForms. However, looking at the variation of mixed Hodge structure~\VarMixed\ for relative three forms we notice that the moduli, $u$, must enter the three-form component of relative cohomology elements. This becomes apparent by taking a derivative, $\partial_u$, of a pure three-form piece appearing in the upper row of the variational diagram~\VarMixed. It yields a two-form contribution in the lower row of the diagram. We readily implement this moduli dependence by representing a closed relative three form, $\rel{\Xi}$, which arises from a pure closed three form, $\Xi$, as
\eqn\RThreeForms{\rel{\Xi}\,=\,(\Xi,0)\,=\,\int {p(x) \log Q \over P^k}\,\Delta \ . }
This definition is now in agreement with the variation of mixed Hodge structure~\VarMixed. Analogously we enhance the two forms~\TwoForms\ to relative two forms, $\rel{\alpha}=(\alpha,0)$,
\eqn\RTwoForms{
  \rel{\alpha}\,=\,
  \int \sum_{m<n} {(a_n\,x_n\,q_m(x) - a_m\,x_m\,q_n(x))\log Q\over P^k}\,
  (-1)^{m+n} dx_1\wedge\ldots\widehat{dx_m}\ldots\widehat{dx_n}\ldots\wedge dx_5 \ , }
and by acting with the de~Rham differential, $d$, we arrive at
\eqn\dRalpha{
  d\rel{\alpha}\,=\,\int\sum_m \left[k{q_m(x) \partial_m P \log Q \over P^{k+1}}
  -{\partial_m q_m(x) \log Q\over P^k} -{q_m(x)\partial_m Q\over P^k Q} \right] \Delta \ . }
Note that due to the introduced $\log Q$-term the relative three form, $d\rel\alpha$, is indeed a relative exact form because it also contains the pullback term~\PBalpha, which is required by the relative cohomology equivalence relation~\RelThreeEquiv. 

Although the definitions of relative forms~\RThreeForms\ and \RTwoForms\ yield the correct equivalence relation~\RelThreeEquiv\ and agrees with the variation of mixed Hodge structure~\VarMixed, the introduction of the $\log Q$-term may seem a little {\it ad hoc}. There is, however, another reason, which suggests the appearance of the $\log Q$-term. The mixed Hodge structure of relative forms, $H^3(Y,V)$, can also be defined as the filtration arising form the hypercohomology of the complexes, $\Omega^*_Y(\log Q)$, \ie the spectral sequence of the hypercohomology degenerates at the term, $E^{p,q}_1 = H^q(Y,\Omega^p_Y(\log Q))$, and abuts to the relative cohomology group, $H^{p+q}(Y,V)$ \refs{\DeligneIII,\VoisinAA}. The forms, $\Omega^*_Y(\log Q)\equiv\Lambda^*\Omega_Y(\log Q)$, are locally generated by the one forms, $\Omega^1(Y)$, and the logarithmic differential, $dQ/Q\equiv d(\log Q)$. Therefore in the decomposition~\HThreeRel\ the $\log Q$-term comes about naturally for the three-form component, corresponding to $p=0$ and $q=3$, as it generates the logarithmic differential, $d(\log Q)$, of the two-form components.

%%%%%%%%%%%%%%%%%%%%%%
\subsec{Open/closed Picard-Fuchs differential equations and flat coordinates}
\subseclab\secPFEq
%%%%%%%%%%%%%%%%%%%%%%
Having developed the techniques to render relative three forms as residue integrals we are now ready to make the connection to the advertised Picard-Fuchs equations. Their solutions are the relative three-form periods~\RelPeriodsTwo, which in turn determine flat coordinates of the open-/closed-string moduli space.

Let us first discuss the system of linear Gauss-Manin differential equations, which controls the mixed Hodge filtration of the relative cohomology, $H^3(Y,V)$, fibered over the open-/closed-string moduli space. We introduce a basis vector, $\rel\pi(z,u)$, of the relative three-form cohomology elements compatible with the mixed Hodge filtrations~\FFilt\ and \WFilt. In practice such a basis is constructed from the unique relative three form, $\rel\Omega(z,u)$, which spans the filtration module, $F^3$,
\eqn\RHolOmega{\rel{\Omega}(z,u)\,=\,(\Omega,0)\,=\,\int {\log Q(u) \over P(z)} \Delta \ . }
Recall that the dependence on the bulk moduli, $z$, arises in the polynomial, $P(z)$, whereas the open-string moduli, $u$, appear in the family of divisors, $Q(u)$. Due to Griffiths transversality we now generate a basis vector, $\rel\pi(z,u)$, by taking consecutive derivatives of the relative three form, $\rel\Omega(z,u)$. Furthermore subsets of this basis span the various filtration modules according to the variational diagram~\VarMixed.

The Gauss-Manin system is the system of the linear differential equations, which expresses infinitesimal  variations of the basis vector, $\rel\pi(z,u)$, with respect to the moduli in terms of a linear combination of the basis elements, $\rel\pi(z,u)$. Hence it reads:
\eqn\VarBasis{\eqalign{
  0\,&\simeq\,\left(\partial_{z_k} - M_k(z,u)\right)\rel\pi(z,u)
     \,\equiv\,\nabla_{z_k} \rel\pi(z,u)  \ , \cr
  0\,&\simeq\,\left(\partial_{u_{\hat k}}- M_{\hat k}(z,u)\right)\rel\pi(z,u)
      \,\equiv\,\nabla_{u_{\hat k}} \rel\pi(z,u)  \ . }}
Here `$\,\simeq\,$' indicates equality on the level of cohomology classes, \ie equality modulo exact relative forms~\RelThreeEquiv. This reflects the fact that in varying the basis vectors, $\rel\pi$, we also modify the representatives of the relative cohomology classes. The indices, $k$ and $\hat k$, label the closed- and open-string moduli respectively. In the next sections we discuss in detail how to employ the residue integrals techniques so as to compute the matrices, $M_k$ and $M_{\hat k}$, explicitly.

The linear Gauss-Manin system~\VarBasis\ gives also rise to the Gauss-Manin connection, $\nabla_{z_k}$ and $\nabla_{u_{\hat k}}$, for the relative three-form mixed Hodge bundle over the open-/closed-string moduli space. The Gauss-Manin connection is actually flat \refs{\LercheCK,\LercheYW}
\eqn\GMFlatness{ 
  [\nabla_{z_k},\nabla_{z_\ell}] \,=\, 
  [\nabla_{u_{\hat k}},\nabla_{u_{\hat\ell}}] \,=\, 
  [\nabla_{z_k},\nabla_{u_{\hat k}}]\,=\,0 \ . }
Analogously to the flat Gauss-Manin connection in the context of $N=2$ special geometry \refs{\CecottiME,\BershadskyCX,\LercheUY}, the flatness of the connection is not a coincidence but a non-trivial and necessary property of the $N=1$ special geometry governing the open/closed chiral ring \refs{\LercheCK,\LercheYW}.

Since the whole relative basis vector, $\rel\pi(z,u)$, is induced from derivatives of the relative three form, $\rel\Omega(z,u)$, we can readily construct from the linear Gauss-Manin system the differential  Picard-Fuchs equations for the relative three form, $\rel\Omega(z,u)$, 
\eqn\PFOmega{ \cL_A \rel\Omega(z,u) \, \simeq \, 0 \ . }
According to the mixed Hodge variation~\VarMixed\ the linear Gauss-Manin system translates into Picard-Fuchs operators, $\cL_A$, that are partial differential operators up to fourth order. As before the differential equations~\PFOmega\ hold on the level of relative cohomology classes. As a consequence the Picard-Fuchs operators also annihilate the relative periods~\RelPeriodsTwo 
\eqn\PFPeriod{ \cL_A \rel\Pi^a(z,u) \,=\,0 \ . }
We should stress that this set of Picard-Fuchs equations is only integrable due to the flatness of the Gauss-Manin connection~\GMFlatness.

The solutions of the open/closed Picard-Fuchs equations~\PFPeriod\ are the relative periods governing the effective superpotential~\WRelPeriod. In particular, as argued in Section~\secFluxBraneSuper, this superpotential comprises both the pure closed-string periods for the RR~three form fluxes and the true relative periods for the brane induced superpotential terms. Thus the bulk periods, $\Pi^\alpha(z)$, are also annihilated by the open-/closed-string Picard-Fuchs operators, $\cL_A$. As a consequence they exhibit the following general structure
\eqn\PFOpStructure{
  \cL_A \,=\, \cL^{\rm bulk}_A(z,\partial_z) + \cL^{\rm bdry}_A(z,u,\partial_z,\partial_u) \ . }
The operators, $\cL^{\rm bulk}_A$, are the Picard-Fuchs operators of the closed-string bulk theory, whereas the operators, $\cL^{\rm bdry}_A$, communicate the connection to the open-string boundary sector. Moreover the latter operators must annihilate the closed-string periods, $\Pi^\alpha(z)$, to ensure that they are indeed solutions to the open-/closed-string Picard-Fuchs equations.

A complete set of relative periods, $\rel\Pi^a$, which solves the integrable Picard-Fuchs system~\PFPeriod, provides for a tool to compute flat coordinates of the open-/closed-string moduli space. By choosing a symplectic basis for the homology group, $H_3(Y,\IZ)$, of three cycles, the closed string periods split into A- and B-periods. Then the flat coordinates, $t$, of the bulk sector are defined by
\eqn\BulkFlatCoord{ t_k(z) \,=\, {\rel\Pi^k(z) \over\rel\Pi^0(z)} \ . }
Here the periods, $\rel\Pi^0(z)$ and $\rel\Pi^k(z)$, constitute the A-periods with respect to the chosen symplectic basis of three cycles. Analogously the open flat coordinates, $\hat t$, arise from an appropriate subset of semi-periods and are defined by
\eqn\OpenFlatCoord{ \hat t_{\hat k}(z,u) \,=\, {\rel\Pi^{\hat k}(z,u)\over \rel\Pi^0(z)} \ . }

As a consequence of $N=1$ special geometry the relative period vector, expressed in terms of flat coordinates, $t$ and $\hat t$, becomes \refs{\LercheCK,\LercheYW}
\eqn\FlatPeriods{ \rel\Pi^a(t,\hat t)\,=\,
   \left(1\,,\ t_k\,,\ \partial_{t_k} \cF(t)\,,\ 2\,\cF(t)-\sum_k t_k\,\partial_{t_k}\cF(t)\,;\
           \hat t_{\hat k}\,,\ W_\ell(t,\hat t)\,,\ * \right) \ . }
The first two entries are the closed-string A-periods encoding the flat coordinates~\BulkFlatCoord. The next two entries correspond to the symplectic dual closed-string B-periods, which, as a consequence of the underlying $N=2$ special geometry, are induced form the closed-string holomorphic $N=2$ prepotential, $\cF(t)$ \StromingerPD. The other periods arise from open-string semi-periods governed by $N=1$ special geometry \refs{\LercheCK,\LercheYW}. Here the first entry gives rise to the open-string flat coordinates~\OpenFlatCoord, whereas the second entry yields the holomorphic D5-brane superpotential components, $W_\ell(t,\hat t)$, associated to the various two cycles labelled by index $\ell$. In general these superpotential components are not integrable to a generating function. This reflects the fact that $N=1$ special geometry is not as constraining as its $N=2$ relative. The remaining semi-periods, which do not allow for an interpretation as flat coordinates or superpotentials, are denoted by `$\,*\,$'. These semi-periods do not appear in refs.~\refs{\LercheCK,\LercheYW} because of the non-compactness of the considered local Calabi-Yau geometries. However, it would be interesting to find a physics interpretation for these semi periods in the context of compact geometries.

Computing flat coordinates, $t$ and $\hat t$, involves a choice of basis, $\rel\Gamma^a$, of the homology group, $H_3(Y,V,\IZ)$. However, at special points in the open-/closed-string moduli space the choice of basis is further restricted such that we can derive an open-/closed-string mirror map. Most prominent is the mirror map at the large complex structure point as it allows to compute instanton corrections to the classical mirror geometry in the topological A-model. Recently, however, it has been demonstrated in local Calabi-Yau geometries that under certain circumstances a mirror map can also be computed at orbifold points in the moduli space \refs{\AganagicWQ,\BouchardYS,\BriniRH,\BouchardGU}, which then allows to determine equivariant invariants in the topological A-model \refs{\CoatesAA,\BayerAA,\BouchardAA}. For our examples discussed in the next sections we  explicitly extract also the latter type of enumerative disk invariants in the context of D5-branes in compact Calabi-Yau manifolds.

%%%%%%%%%%%%%%%%%%%%%%
\newsec{D5-branes in the degree eight hypersurface in ${\bf\IW\IP^{4}_{(1,1,1,1,4)}/(\IZ_8)^2\times\IZ_2}$}
\seclab\secExampleOne
%%%%%%%%%%%%%%%%%%%%%%
In this section we apply the developed tools to our first example. We consider the Calabi-Yau threefold, which arises as the mirror of the family of degree eight hypersurfaces in the weighted projective space, $\IW\IP^{4}_{(1,1,1,1,4)}$. The analyzed D-brane geometry in this Calabi-Yau space allows us to derive  open/closed Picard-Fuchs equations. Their solutions yield the effective superpotential in flat coordinates. From this superpotential we extract a certain domain wall tension, which, for this example, is calculated by different means in ref.~\refs{\KnappUW,\KreflSJ} in agreement with our results. 

%%%%%%%%%%%%%%%%%%%%%%
\subsec{The Calabi-Yau and D-brane geometry}
%%%%%%%%%%%%%%%%%%%%%%
The first task is to specify the bulk Calabi-Yau geometry. The degree eight hypersurfaces in the weighted projective space, $\IW\IP^{4}_{(1,1,1,1,4)}$, give rise to a family of Calabi-Yau threefolds with one K\"ahler modulus and $149$ complex structure moduli \KlemmTX. Here we are mainly interested in its mirror family of Calabi-Yau threefolds, $Y$, depending on one complex structure modulus and $149$ K\"ahler moduli. Applying the standard Greene-Plesser construction \GreeneUD, we realize the mirror (in its singular limit) as the degree eight hypersurface
\eqn\POne{ P(\psi) \,=\, x_1^8 + x_2^8 + x_3^8 + x_4^8 + x_5^2 - 8\,\psi\,x_1 x_2 x_3 x_4 x_5 \ , } 
in the $(\IZ_8)^2\times\IZ_2$-orbifold of the weighted projective space, $\IW\IP^{4}_{(1,1,1,1,4)}$. The parameter, $\psi$, is the algebraic complex structure modulus of the Calabi-Yau threefold, $Y$. The orbifold group is generated by\foot{Naively these generators give rise to the group, $(\IZ_8)^3$. However, a $\IZ_4$-subgroup acts trivially due to quasi-homogeneous identifications of the projective coordinates \KlemmTX.} 
\eqn\OrbGen{ g_1=(1,0,0,7,0) \ , \quad g_2=(0,1,0,7,0) \ , \quad g_3=(0,0,1,7,0) \ , }
acting on the weighted projective coordinates, \eg for the generator, $g_1$, we get
\eqn\GenAction{
  g_1: \ [\,x_1 : x_2 : x_3 : x_4 : x_5\,] \mapsto [\,\eta\,x_1 : x_2 : x_3 : \eta^7x_4 : x_5\,] \ , 
  \quad \eta\equiv e^{2\pi i/8} \ . }
Resolving the orbifold singularities by standard toric techniques one obtains a smooth family of Calabi-Yau threefolds depending on  $149$ K\"ahler moduli.

The next step is to introduce the divisor, $V$, which determines the D5-brane contents in our example. The divisor, $V$, is defined by the polynomial
\eqn\PolDivOne{ Q(\phi) \,=\, x_5-\phi\,x_1 x_2 x_3 x_4 \ . }
Here $\phi$ is the algebraic open-string modulus. Note that this constitutes the most general polynomial of degree four invariant with respect to the $(\IZ_8)^2\times \IZ_2$-orbifold group.

The family of branes defined by the divisor, $Q$, are directly related to the D5-branes in the Calabi-Yau, $Y$, discussed in refs.~\refs{\KnappUW,\KreflSJ}. There the relevant D5-branes wrap the holomorphic two cycles, $C_\pm$, which are respectively given in the ambient weighted projective space by\foot{In ref.~\KnappUW\ the Calabi-Yau threefold, $Y$, is given by the degree eight hypersurface polynomial, $\tilde P(\tilde\psi)=x_1^8 + x_2^8 + x_3^8 + x_4^8 + x_5^2 - 4\tilde\psi\,x_1^2 x_2^2 x_3^2 x_4^2$, and the holomorphic two cycles, $C_\pm$, are defined as $C_\pm=\{ x_5 \pm 2\sqrt{\tilde\psi}\,x_1 x_2 x_3 x_4=0\, , x_1 + \mu\,x_2 = 0\, , x_3+\nu\,x_4 = 0\, ,\mu^8=\nu^8=-1\}$. These definitions translate to our conventions if we identify the algebraic closed string moduli as $\tilde\psi \equiv 4\psi^2$ and change the weighted projective coordinates according to $x_5 \rightarrow x_5 - 4 \psi x_1 x_2 x_3 x_4$.} 
\eqn\PolC{\eqalign{ 
  C_+\,&=\,\left\{ x_5 = 0\, , x_1 + \mu\,x_2 = 0\, , x_3+\nu\,x_4 = 0\, ,
      \mu^8=\nu^8=-1 \right\} \subset Y \ , \cr
  C_-\,&=\,  \left\{ x_5 - 8\,\psi\,x_1 x_2 x_3 x_4=0\, , x_1 + \mu\,x_2 = 0\, , x_3+\nu\,x_4 = 0\, ,
      \mu^8=\nu^8=-1 \right\} \subset Y \ .}}
We make the important observation that the curves, $C_\pm$, lie in the divisors given in terms of the polynomial, $Q(\phi)$, evaluated at the two critical points,
\eqn\CPoints{ \phi_+ \,=\, 0 \ , \quad \phi_- \,=\, 8\,\psi \ . }
Therefore we claim that the family of divisors, $V$, gives rise to the relative period integrals, which at the critical points describe the D5-branes, $C_\pm$. This claim is supported by the fact that the spectrum of the two D5-branes, $C_\pm$, consists of one (obstructed) open-string modulus. Moreover, this modulus enters a cubic superpotential, and the two brane configurations, $C_\pm$, emerge at the two critical points of this superpotential \KnappUW. As we go along and compute the effective superpotential in flat coordinates we further substantiate this picture and make this correspondence more precise.

Before we conclude this section we establish that the open/closed moduli space parametrized by the algebraic moduli, $\psi$ and $\phi$, exhibits a $\IZ_8\times\IZ_2$-symmetry. The generator of the $\IZ_8$-group acts on the algebraic moduli as
\eqn\DiscreteEight{ \pmatrix{ \psi \cr \phi} \mapsto \eta \pmatrix{ \psi \cr \phi } \ , \quad \eta\equiv e^{2\pi i/8} \ .}
This is indeed a symmetry as its action on the polynomials, $P(\psi)$ and $Q(\phi)$, is readily compensated by the projective coordinate transformation, $x_1\rightarrow \eta^{-1}x_1$. The $\IZ_2$-symmetry is generated by
\eqn\DiscreteTwo{ \pmatrix{ \psi \cr \phi} \mapsto \pmatrix{ \psi \cr 8\psi-\phi } \ ,}
and its action on the polynomials, $P(\psi)$ and $Q(\phi)$, is balanced by the projective coordinate transformation, $[x_1: x_2:x_3:x_4:x_5] \rightarrow [-x_1: x_2:x_3:x_4:x_5-8\,\psi\,x_1x_2x_3x_4]$. Note that the $\IZ_2$~symmetry exchanges the two critical points, $\phi_\pm$, and hence maps a D5-brane wrapped on the holomorphic two cycle,  $C_+$, to a D5-brane wrapped on the holomorphic two cycle, $C_-$, and {\it vice versa}. Thus altogether the genuine open/closed moduli space is really a $\IZ_8\times\IZ_2$-orbifold of the covering space parametrized by the algebraic variables, $\psi$ and $\phi$. The corresponding orbifold singularities emerge at the fixed point loci, $\psi=0$ and $\phi=4\psi$, of the $\IZ_8$- and $\IZ_2$-group action~\DiscreteEight\ and \DiscreteTwo.

%%%%%%%%%%%%%%%%%%%%%%
\subsec{The linear Gauss-Manin system of differntial equations}
%%%%%%%%%%%%%%%%%%%%%%
In this section we explicitly construct the linear Gauss-Manin system of differential equations for the D5-brane geometry alluded to in the previous section. By taking derivatives of the relative holomorphic three form,
\eqn\holthree{\rel{\Omega}(\psi,\phi)\,=\,\int {\log Q(\phi)\over P(\psi)}\, \Delta\ ,}
defined in terms of the polynomials~\POne\ and \PolDivOne, we generate due to Griffiths transversality and according to the variational diagram~\VarMixed\ a basis for the relative cohomology group, $H^3(Y,V)$. A convenient basis turns out to be
\eqn\BasisOne{
  \rel\pi \equiv
  \left( \pi_{3,3} , \pi_{2,3} , \pi_{1,3} , \pi_{0,3} , \pi_{2,4} , \pi_{1,4} , \pi_{0,4} \right) =
  \left( \rel\Omega , \partial_\psi\rel\Omega , \partial_\psi^2\rel\Omega , \partial_\psi^3\rel\Omega ,
          \partial_\phi\rel\Omega , \partial_\psi\partial_\phi\rel\Omega ,
          \partial_\psi^2\partial_\phi\rel\Omega \right) \ . }
All these basis elements are represented by relative three-form residue integrals 
\eqn\ResBasisOne{\eqalign{
  \pi_{3-k,3}&=k!\,8^{k}\int {(x_{1}x_{2}x_{3}x_{4}x_{5})^{k}\over P(\psi)^{k+1}}\,\log Q(\phi)\,\Delta\ ,\qquad k\,=\,0,1,2,3\ ,\cr
  \pi_{2-k,4}&=-k!\,8^{k}\int {(x_{1}x_{2}x_{3}x_{4})^{k+1}x_{5}^{k}\over P(\psi)^{1+k}Q(\phi)}\,\Delta\ ,\qquad k\,=\,0,1,2\ . }}
Note that the structure of the chosen basis is in accord with the mixed Hodge filtration defined in Section~\secMixedHodge. In particular the bases for the decreasing filtration modules, $F^p$, for $p=3,2,1,0$ are given by $\pi_{3,3},\ldots,\pi_{p,3},\pi_{2,4},\ldots,\pi_{p,4}$, whereas the increasing weight filtration modules, $W_3$ and $W_4$, are spanned by the basis vectors, $(\pi_{\ell,3})_{\ell=3,\ldots,0}$ and $\rel\pi$, respectively.

The next task is to determine the linear Gauss-Manin system of differential equations~\VarBasis\ in terms of the above defined basis. Therefore we expand the vectors, $\partial_\psi\rel\pi$ and $\partial_\phi\rel\pi$, into the defined basis elements, $\pi_{k,l}$. This procedure is trivial for some of the differentiated basis elements, namely directly from the definition of the basis vector~\BasisOne\ we can read off the following relations
\eqn\TrivRelations{\eqalign{
  \partial_\psi \pi_{k,4} \,&=\, \pi_{k+1,4} \quad {\rm for}\quad k=2,1 \ , \cr
  \partial_\psi \pi_{k,3} \,&=\, \pi_{k+1,3} \quad {\rm for}\quad k=3,2,1 \ , \cr
  \partial_\phi \pi_{k,3} \,&=\, \pi_{k+1,4} \quad {\rm for}\quad k=3,2,1 \ .}}
However, in order to find the expansions for the derivatives of the remaining basis elements we are required to employ the whole machinery of relative three-form residue integrals developed in Section~\secResRelThree. That is to say we need to add in a systematic manner appropriate exact relative three forms~\dbeta\ and \dRalpha\ so as to establish the correct expansions. The performed reduction procedure constitutes a generalization of the Griffiths-Dwork algorithm in the context of relative form residue integrals. While the technical details of these long and tedious calculations are relegated to Appendix~A, we simply collect the results here. For the remaining derivatives acting on the basis elements, $\pi_{k,4}$, we find 
\eqn\DPhiPiFour{\eqalign{
  \partial_\phi\pi_{2,4}\,&\simeq\, {4\psi-\phi\over 4\phi}\,\pi_{1,4}  \ , \cr
  \partial_\phi\pi_{1,4}\,&\simeq\, {4\psi-\phi \over 4\phi}\,\pi_{0,4}+{1\over \phi}\,\pi_{1,4} \ , \cr
  \partial_\phi\pi_{0,4}\,&\simeq\, {R_{1}\over D_2}\,\pi_{2,4}
      +{R_{2}\over D_2}\,\pi_{1,4}+{R_{3}\over\phi D_2}\,\pi_{0,4} \ . }}
As before `$\,\simeq\,$' indicates that these equations hold on the level of cohomology classes. The polynomials, $R_{1}$, $R_{2}$, and $R_{3}$, are given by
\eqn\Rpolys{\eqalign{
  R_{1}\,&=\,-128\,\phi ^3 (\phi -8 \psi ) (\phi -4 \psi ) \ ,\cr
  R_{2}\,&=\,112\,\phi ^3 (\phi -8 \psi )^2 (\phi -4 \psi )\ ,\cr
  R_{3}\,&=\,-2\left(5 \phi ^8-136 \psi  \phi ^7+1344 \psi ^2 \phi^6-5632 \psi ^3 \phi ^5+8192 \psi ^4 \phi ^4+256\right)\ ,}}
and we further introduce the two discriminants
\eqn\discrimtwo{
  D_1\,=\,1-(2\psi)^8 \ , \qquad
  D_2\,=\,\prod_{\ell=1}^4 \left(\phi(\phi-8 \psi) - 4\,e^{\pi i\ell\over 2} \right) \ .}
As the basis elements~\BasisOne\ arise from the variation of the relative three form, $\rel\Omega$, it implies $\partial_\psi\pi_{0,4}=\partial_\phi\pi_{0,3}$, and we find
\eqn\DPsiPiFour{
  \partial_\psi\pi_{0,4}\,=\,\partial_\phi\pi_{0,3}
  \,\simeq\, {S_1\over D_2}\,\pi_{2,4}+{S_{2}\over D_2}\,\pi_{1,4}+{S_{3}\over  D_2}\,\pi_{0,4}  \ . }
Here the polynomials, $S_{1}$, $S_{2}$, and $S_{3}$, are given by
\eqn\Spoly{S_{1}\,=\, 512\, \phi ^4 (\phi -8 \psi )\ ,\quad
S_{2}\,=\,-448\, \phi ^4 (\phi -8 \psi )^2\ ,\quad
S_{3}\,=\, 48\, \phi ^4 (\phi -8 \psi )^3\ .}
We observe that the expansions of the derivatives acting on the elements, $\pi_{k,4}$, do not involve the basis elements, $\pi_{k,3}$. This manifests in this example the existence of the variational sub-system~\VarSubSys\ within the variation of mixed Hodge structure~\VarMixed. Finally it remains to expand the element, $\partial_\psi\pi_{0,3}$, for which we obtain
\eqn\DPsiPiThree{\eqalign{
  \partial_\psi\pi_{0,3}\,\simeq&\, {256\psi^{4}\over D_1}\,\pi_{3,3}+{15(1+(2\psi)^{8})\over \psi^{3}D_1}\,\pi_{2,3}-{5(3-1280\psi^{8})\over \psi^{2}D_1}\,\pi_{1,3}+{2(3+1280\psi^{8})\over \psi D_1}\,\pi_{0,3}\cr
  &+{T_{1}\over\psi^3D_{1}D_2}\,\pi_{2,4}+{T_{2}\over \phi\psi^2D_1D_{2}}\,\pi_{1,4}+{T_{3}\over \phi^2\psi D_1 D_2}\,\pi_{0,4}\ ,}}
in terms of the polynomials\foot{Note that the polynomials, $T_1$ to $T_3$, are actually finite for $\phi\rightarrow 4\psi$.}
\eqn\TSpoly{{\eqalign{
   T_{1}\,=&\,{4\over\phi-4\psi}\Big(\left(256\, \phi  \psi^8+448\,\phi ^2 \psi^7-60\,\psi +15\,\phi\right) 
       D_2 -\psi^3\phi\,S_1 D_1  \Big) \ ,\cr
   T_{2}\,=&\,{4\over\phi-4\psi}\Big(\left(1792 \phi ^2 \psi ^8+1792 \phi ^3 \psi ^7
         -112 \phi ^4 \psi ^6+48 \psi ^2+48 \phi  \psi -15 \phi ^2\right) D_2  \cr 
          & \qquad\qquad\qquad -\psi^2\phi^2S_2 D_1  \Big) \cr
   T_{3}\,=&\,{4\over\phi-4\psi}\Big(2\left(768 \phi ^3 \psi ^8+352 \phi ^4 \psi ^7
          -48 \phi ^5 \psi ^6+2 \phi ^6 \psi ^5-32 \psi ^3\right. \cr
          & \qquad\qquad\qquad \left.-16 \phi  \psi ^2-6 \phi
   ^2 \psi +3 \phi ^3\right) D_2 -\psi\,\phi^3S_3 D_1  \Big)  \ . }}}
All the relations~\TrivRelations\ to \DPsiPiThree\ are now summarized in the connection matrices, $M_\phi$ and $M_\psi$, of the linear Gauss-Manin system~\VarBasis. They read
\eqn\MPsiOne{
  M_\psi\,=\pmatrix{
     0 & 1 & 0 & 0 & 0 & 0 & 0 \cr
     0 & 0 & 1 & 0 & 0 & 0 & 0 \cr
     0 & 0 & 0 & 1 & 0 & 0 & 0 \cr
     {256 \psi ^4 \over D_1} & {3840 \psi ^8+15\over\psi^3 D_1} & -{15-6400 \psi ^8\over\psi ^2 D_1} & 
          {2560 \psi ^8+6\over\psi D_1} & {T_1\over\psi^3 D_1 D_2} &
        {T_2\over \phi \psi ^2 D_1 D_2}& {T_3\over\phi ^2 \psi D_1 D_2} \cr
 0 & 0 & 0 & 0 & 0 & 1 & 0 \cr
 0 & 0 & 0 & 0 & 0 & 0 & 1 \cr
 0 & 0 & 0 & 0 & {S_1\over D_2} & {S_2\over D_2} & {S_3\over D_2} }\ , }
and
\eqn\MPhiOne{
  M_\phi\,=\pmatrix{
    0 & 0 & 0 & 0 & 1 & 0 & 0 \cr
    0 & 0 & 0 & 0 & 0 & 1 & 0 \cr
    0 & 0 & 0 & 0 & 0 & 0 & 1 \cr
    0 & 0 & 0 & 0 & {S_1\over D_2} & {S_2\over D_2} & {S_3\over D_2} \cr
    0 & 0 & 0 & 0 & 0 & {\psi\over\phi }-{1\over 4} & 0 \cr
    0 & 0 & 0 & 0 & 0 & {1\over\phi} & {\psi\over\phi}-{1\over 4} \cr
    0 & 0 & 0 & 0 & {R_1\over D_2} & {R_2\over D_2} & {R_3\over\phi  D_2}  }\ . }
The discriminant, $D_1$, corresponds to the (mirror) conifold locus appearing already in the pure bulk geometry. At the zero locus of the discriminant, $D_2$, in the moduli space, the intersection of the two polynomials, $P$ and $Q$, fails to be transversal at some points in the embedding space, $\IW\IP^4_{(1,1,1,1,4)}$, and as a consequence at these points the D5-brane divisor, $V$, becomes singular. Let us also remark that the matrices~\MPhiOne\ reveal a block diagonal structure. As indicated before this demonstrates for this example that the variation of mixed Hodge structure~\VarMixed\ contains the sub-system~\VarSubSys.

Let us now focus on the Gauss-Manin connection, $\nabla_\phi\equiv\partial_\phi-M_\phi$ and $\nabla_\psi\equiv\partial_\psi-M_\psi$, of the open-/closed-string moduli space. As discussed in detail in Section~\secPFEq\ the underlying $N=1$ special geometry requires the Gauss-Manin connection to be flat. For our example, which involves only two moduli, $\psi$ and $\phi$, the conditions~\GMFlatness\ reduce to a single equation
\eqn\integcon{\partial_{\phi}M_{\psi}-\partial_{\psi}M_{\phi}+[M_{\psi},M_{\phi}]=0\ ,}
which is indeed satisfied for the constructed connection matrices~\MPsiOne\ and \MPhiOne. We should stress again that the flatness of this connection is a non-trivial condition imposed on the linear Gauss-Manin system. It insures that the associated system of Picard-Fuchs equations, which we examine in the next sections, are integrable. Furthermore the fulfilled integrability condition~\integcon\ serves also as a non-trivial check on our method of realizing the variation of mixed Hodge structure in terms of relative three-form residue integrals.

%%%%%%%%%%%%%%%%%%%%%%
\subsec{Relative periods in the vicinity of the orbifold point}
%%%%%%%%%%%%%%%%%%%%%%
First we want to solve the linear Gauss-Manin system parametrized by the algebraic moduli, $\psi$ and $\phi$, in the vicinity of the fixed point locus of the symmetry transformations~\DiscreteEight\ and \DiscreteTwo, \ie in the vicinity of the $\IZ_8\times\IZ_2$ orbifold singularity of the open/closed string moduli space located at $\psi=0$ and $\phi=4\psi=0$. From the linear Gauss-Manin system characterized by the two $7\times 7$ matrices~\MPhiOne\ and \MPsiOne\ we extract the following Picard-Fuchs operators~\PFOmega
\eqn\PFOne{
   \cL_1\,=\, \widetilde\cL_1\, \partial_\phi \ , \quad
   \cL_2\,=\, \widetilde\cL_2\, \partial_\phi \ , \quad
   \cL_3\,=\, \cL^{\rm bulk} + \cL^{\rm bdry} \ .}
The differential operators, $\widetilde\cL_1$ and $\widetilde\cL_2$, read
\eqn\PFSubOne{\eqalign{
   \widetilde\cL_1 \,&=\,(4\psi-\phi)\,\theta_\psi-4\psi\,\theta_\phi \ , \cr
   \widetilde\cL_2 \,&=\,\partial^{2}_{\psi}\partial_{\phi}-{48\over D_{2}}\,\phi ^4 (\phi -8 \psi )^3\,\partial^{2}_{\psi}+{448\over D_{2}}\, \phi ^4 (\phi -8 \psi )^2\,\partial_{\psi}-{512\over D_{2}}\, \phi ^4 (\phi -8 \psi )\ , }}
and the operators, $\cL^{\rm bulk}$ and $\cL^{\rm bdry}$, are
\eqn\PFWholeOne{\eqalign{
   \cL^{\rm bulk} \,&=\,\theta_\psi(\theta_\psi-2)(\theta_\psi-4)(\theta_\psi-6)-(2\psi)^8 (\theta_\psi+1)^4 \ , \cr
   \cL^{\rm bdry} \,&=\,\psi^4D_1\partial^{4}_{\psi}-{\psi^3\,T_3\over\phi ^2D_2}\,\partial^{2}_{\psi}\partial_{\phi}-{\psi^2\,T_2\over \phi D_2}\,\partial_{\psi}\partial_{\phi}-{\psi\,T_1\over D_2}\,\partial_{\phi}\ . }}
in terms of the logarithmic derivatives, $\theta_\psi\equiv\psi\,\partial_\psi$ and $\theta_\phi\equiv\phi\,\partial_\phi$. Note that the linear Gauss-Manin system is equivalent to two independent partial differential operators of up to degree four. Thus one of the operators, $\cL_1$ or $\cL_2$, is redundant. However, in order to find solutions to the partial differential equations associated to these rather complicated operators we take advantage of the variational sub-system~\VarSubSys\ governed by the two Picard-Fuchs operators, $\widetilde\cL_1$ and $\widetilde\cL_2$. 

Before we investigate the sub-system~\VarSubSys\ we observe that the Picard-Fuchs operators~\PFOne\ exhibit indeed the structure advocated in eq.~\PFOpStructure. Namely, as discussed in Section~\secPFEq\ the bulk periods, $\Pi^\alpha(\psi)$, determined by the hypergeometric differential equation, $\cL^{\rm bulk}\Pi^\alpha(\psi)=0$, form a subset of solutions to the open/closed Picard-Fuchs equations. Hence these periods are given in terms of hypergeometric functions \RainvilleBook
\eqn\BulkSolOne{\eqalign{
  \rel\Pi^0(\psi)\equiv\Pi^0(\psi)\,&=\,
    \HGF{4}{3}\left({1\over8},{1\over8},{1\over8},{1\over8};{1\over4},{2\over4},{3\over4};(2\psi)^8\right) \ , \cr
   \rel\Pi^1(\psi)\equiv\Pi^1(\psi)\,&=\, (2\psi)^2\,
    \HGF{4}{3}\left({3\over8},{3\over8},{3\over8},{3\over8};{2\over4},{3\over4},{5\over4};(2\psi)^8\right) \ , \cr
   \rel\Pi^2(\psi)\equiv\Pi^2(\psi)\,&=\, {(2\psi)^4\over 2!}\,
    \HGF{4}{3}\left({5\over8},{5\over8},{5\over8},{5\over8};{3\over4},{5\over4},{6\over4};(2\psi)^8\right) \ , \cr
   \rel\Pi^3(\psi)\equiv\Pi^3(\psi)\,&=\, -{(2\psi)^6\over 3!}\,
    \HGF{4}{3}\left({7\over8},{7\over8},{7\over8},{7\over8};{5\over4},{6\over4},{7\over4};(2\psi)^8\right) \ , }}
which enjoy for $|\psi|<{1\over 2}$ the convergent expansion
\eqn\BulkSolOneExpansion{ \rel\Pi^\alpha(\psi)\,=\,{c_{\alpha}\over \alpha!\,2^{2\alpha}\,\Gamma\big({2\alpha+1\over 8}\big)^4}\,\sum_{k=0}^{\infty}\,{\Gamma^{4}\big(k+{2\alpha+1\over 8}\big)\over \Gamma(4k+2\alpha+1)}\, (4\psi)^{2(4n+\alpha)}\ , \quad \alpha=0,1,2,3 \ , }
where the constants, $c_{\alpha}$, are all one except for $c_{3}=-1$.

The next task is to determine the solutions of the sub-system described by the two differential operators, $\widetilde\cL_1$ and $\widetilde\cL_2$. We proceed in two steps. First we notice that a general solution to the differential operator, $\widetilde\cL_1$, is given by
\eqn\SolLOne{ \chi(\psi,\phi) \,\equiv\, \chi(u)  \quad {\rm with} \quad u\equiv\phi(\phi-8\psi) \ . }
Second we act with the differential operator, $\widetilde\cL_2$, on the ansatz~\SolLOne\ and obtain an ordinary differential equation for the function, $\chi(u)$,
\eqn\SolLTwo{
  \left[ \theta_u (\theta_u-1) (\theta_u-2) - \left(u\over4\right)^4 (\theta_u-1)^3\right] \chi(u) \,=\,0 \ . }
This is another hypergeometric differential equation with the three linearly independent solutions
\eqn\SolChi{\eqalign{
  \chi^4(u) & = \HGF{3}{2}\left({1\over4},{1\over4},{1\over4};{2\over4},{3\over4}; \left(u\over4\right)^4\right)  \ , \cr
  \chi^5(u) & = \left({u\over4}\right) \, 
    \HGF{3}{2}\left({2\over4},{2\over4},{2\over4};{3\over4},{5\over4}; \left(u\over4\right)^4 \right) \ , \cr
  \chi^6(u) & = \left({u\over4}\right)^2 \,
     \HGF{3}{2}\left({3\over4},{3\over4},{3\over4};{5\over4},{6\over4}; \left(u\over4\right)^4 \right)  \ . }}
The power series of these hypergeometric functions read
\eqn\SolChiExpand{
  \chi^{4+\ell}(u)\,=\,{1\over 4^\ell\,\Gamma\big({1+\ell\over 4}\big)^{4}}\,\sum_{k=0}^{\infty}\,{\Gamma^{4}\big(k+{1+\ell\over 4}\big)\over \Gamma(4k+\ell+1)}\, u^{4k+\ell}\ ,\qquad \ell\,=\,0,1,2\ , }
with radius of convergence~$|u|<4$.

From the solutions~\SolChi\ of the sub-system we construct three additional relative periods, $\rel\Pi^{\hat\alpha}(\psi,\phi)$, which according to the structure of the operators, $\cL_1$ and $\cL_2$, in eq.~\PFOne\ are integrals of the relation 
\eqn\PeriodCondOne{ \partial_\phi\rel\Pi^{\hat\alpha}(\psi,\phi)\,=\, \chi^{\hat\alpha}\left(\phi(\phi-8\psi)\right)  \ . }
The integrals of this equation allow for integration constants, $C^{\hat\alpha}(\psi)$, which have to be chosen such that the integrals are annihilated by the third Picard-Fuchs operator, $\cL_3$. Instead of solving yet another differential equation for $C^{\hat\alpha}(\psi)$, we use the $\IZ_2$~symmetry~\DiscreteTwo\ so as to determine the integration constants. Due to the symmetry the relative periods split into symmetric and anti-symmetric solutions with respect to the $\IZ_2$~action~\DiscreteTwo. The functions~\SolLOne\ are symmetric and, therefore, up to a symmetric bulk period~\BulkSolOne, the integrals must be anti-symmetric in order to be solutions to the whole Picard-Fuchs system, \ie the relative periods, $\rel\Pi^{\hat\alpha}(\psi,\phi)$, vanish at the $\IZ_2$-fixed point locus, $(\psi,\phi)=(\psi,4\psi)$. Hence altogether we arrive at the relative periods
\eqn\PureRelPeriodsIntOne{
  \rel\Pi^{\hat\alpha}(\psi,\phi)\,=\,\int_{4\psi}^\phi \chi^{\hat\alpha}\left(\zeta(\zeta-8\psi)\right)\,d\zeta \ , }
which are indeed annihilated by the Picard-Fuchs operator, $\cL_3$. They give rise to the power series
\eqn\PureRelPeriodsOne{\eqalign{
  \rel\Pi^4(\psi,\phi)\,&=\,\sum_{k=0}^{+\infty} \sum_{n=0}^{4k}
  {\Gamma(1/4+k)^4 \over \Gamma(1/4)^4(4k-n)!\,n!\,(2n+1)}\,(-1)^n\,(\phi-4\psi)^{2n+1}(4\psi)^{8k-2n} \ , \cr
  \rel\Pi^5(\psi,\phi)\,&=\,\sum_{k=0}^{+\infty} \sum_{n=0}^{4k+1}
  {\Gamma(1/2+k)^4 \over \pi^2 (4k+1-n)!\,n!\,(2n+1)}\,(-1)^{n+1}\,(\phi-4\psi)^{2n+1}(4\psi)^{8k+2-2n} \ , \cr
  \rel\Pi^6(\psi,\phi)\,&=\,\sum_{k=0}^{+\infty} \sum_{n=0}^{4k+2}
  {2\,\Gamma(3/4+k)^4 \over\Gamma(3/4)^4(4k+2-n)!\,n!\,(2n+1)}\,(-1)^n\,(\phi-4\psi)^{2n+1}(4\psi)^{8k+4-2n} \ , }}
convergent for $|\phi(\phi-8\psi)|<4$.

In summary the relative periods, $\rel\Pi^a(\psi,\phi),\,a=0,\ldots,6$, of eqs.~\BulkSolOne\ and \PureRelPeriodsOne\ constitute a complete set of solutions to the open/closed Picard-Fuchs equations in the vicinity of the $\IZ_8\times\IZ_2$~orbifold point.

%%%%%%%%%%%%%%%%%%%%%%
\subsec{Effective superpotential and domain wall tension}
%%%%%%%%%%%%%%%%%%%%%%
The next step is to investigate the three-form relative periods computed in the previous section. The reason for analyzing the periods in the vicinity of the $\IZ_8\times\IZ_2$~orbifold point is twofold. First of all we work in a regime containing the two critical points~\CPoints\ simultaneously. Therefore we can directly extract the domain wall tension between two D5-branes wrapping the cycles, $C_\pm$, and compare to the results obtained in refs.~\refs{\KnappUW,\KreflSJ}. Second we are able to extract equivariant orbifold invariants along the lines of refs.~\refs{\BouchardYS,\BouchardGU}.

First we want to determine the flat coordinate, $t$, and the holomorphic prepotential, $\cF$, of the closed-string sector, that is to say we first focus on the first four entries of the relative period vector~\FlatPeriods. In order to define a symplectic basis for the homology group, $H_3(Y,\IZ)$, which singles out an unambiguous choice of flat closed-string periods, we impose the following two criteria. We demand that the flat periods are not singular at the orbifold locus, $\psi=0$, and we require that the flat periods form one-dimensional irreducible representations with respect to the $\IZ_8$-monodromy group action~\DiscreteEight, \ie the transformation, $\psi\rightarrow \eta\,\psi$, induces on flat periods a phase rotation, $\eta^m$, for some integer~$m$.

The physics motivation for the former requirement reflects the fact that, although there is a $\IZ_8$-orbifold singularity in the complex structure moduli space at $\psi=0$, the Calabi-Yau, $Y$, itself is smooth. Therefore the flat periods should also be regular at the orbifold locus. The second criterion comes about as follows. The closed-string conformal field theory of the Calabi-Yau, $Y$, at the orbifold point is captured by a Landau-Ginzburg $\IZ_8$-orbifold theory \WittenYC, whose chiral multiplets fall into $\IZ_8$~representations \IntriligatorUA. Since the holomorphic prepotential together with the flat coordinates encode the chiral ring of this conformal field theory \refs{\LercheUY,\DijkgraafDJ,\StromingerPD}, it is natural to also arrange the flat periods and hence the resulting chiral ring structure constants into $\IZ_8$~representations. 

For our example the discussed two requirements pin down uniquely (up to two numerical constants) the closed-string flat periods~\FlatPeriods, and for the calculated periods~\BulkSolOne\ we derive up to an overall numerical constant the closed-string flat coordinate
\eqn\FlatBulkt{ t(\psi)\,=\,{\rel\Pi^1(\psi)\over\rel\Pi^0(\psi)} \ , }
which yields the closed-string mirror map at the $\IZ_8$-orbifold locus. The first few terms in the expansion explicitly read\foot{Note that the flat coordinate is a function of $\psi^2$. Thus it is convenient to also state the mirror map as a function, $\psi^2(t)$.}
\eqn\MirrorClosedOne{ \psi^2(t)\,=\, 
  {1\over4}t-{19\over1920}t^5-{541\over516\,096}t^9-{2\,177\,327\over7\,084\,965\,888} t^{13}+\ldots \ . }
From the remaining periods we deduce the holomorphic prepotential, $\cF$, which by applying the stated criteria is only ambiguous up to a second numerical factor, and it becomes in terms of the flat coordinate, $t$,  
\eqn\FlatFOne{
   \cF(t)\,=\,{1\over2}\left.{\rel\Pi^3(\psi)\over\rel\Pi^0(\psi)}\right|_{\psi=\psi(t)}
     +\,{t\over2}\left.{\rel\Pi^2(\psi)\over\rel\Pi^0(\psi)}\right|_{\psi=\psi(t)} \ . }
The explicit expansion yields
\eqn\FlatFOneExp{
  \cF(t)\,=\,
     {1\over3!}t^3+{77\over7!\cdot 8}t^7+{161\,071\over 11!\cdot 16} t^{11}
    +{20\,606\,066\,649\over15!\cdot 256} t^{15}
    +{626\,507\,087\,510\,997\over19!\cdot 256} t^{19} +\ldots \ . }
Note that, up to the two undetermined overall numerical scales, this expansion contains the genus zero orbifold invariants of the compact hypersurface, $Y$. It would be interesting to directly establish these invariants by a localization computation in the topological A-model. Such a computation fixes also the mentioned normalization ambiguities.

Next we turn to the open-string sector. Analogously to the closed-string sector we demand that the open-string semi-periods come also in one-dimensional representations of the $\IZ_8$-orbifold group action~\DiscreteEight. Since this condition is already fulfilled for the semi-periods, $\rel\Pi^\alpha$, stated in eq.~\PureRelPeriodsOne, the remaining task is to identify semi-periods for the flat coordinates and the superpotential respectively.

In order to yield a good coordinate system in the vicinity of the orbifold locus the open-string flat coordinate, $\hat t$, must not vanish identically along the $\IZ_8$-orbifold locus, $\psi=0$. This latter requirement, however, is only fulfilled by the semi-period, $\rel\Pi^4(\psi,\phi)$. Therefore, up to a numerical factor, the open-string flat coordinate has to be
\eqn\FlatOpent{ \hat t(\psi,\phi)\,=\,{\rel\Pi^4(\psi,\phi)\over\rel\Pi^0(\psi)} \ . }
From the flat coordinate, $\hat t$, we compute recursively the expansion of the open-mirror map, whose first few terms are given by
\eqn\MirrorOpenOne{\eqalign{ \phi(t,\hat t)\,=\,
    %% degree 1   
     &4\,\psi(t)+\hat t
    %% degree 9 
    -{5\over128}t^4\,{\hat t}+{1\over72}t^3\,{\hat t}^3-{1\over320}t^2\,{\hat t}^5
    +{1\over2688}t\,{\hat t}^7-{1\over55\,296}\,{\hat t}^9 \cr
    %% degree 17
    &\quad
    -{72\,799\over10\,321\,920}t^8\,{\hat t}+{6269\over967\,680}t^7\,{\hat t}^3
    -{21\,337\over5\,529\,600}t^6\,{\hat t}^5\cr
    &\quad\quad
    +{2509\over1\,720\,320}t^5\,{\hat t}^7-{10\,001\over27\,525\,120}t^4\,{\hat t}^9
    +{30\,707\over510\,935\,040}t^3\,{\hat t}^{11}\cr
    &\quad\quad\quad
    -{10\,291\over1\,610\,219\,520}t^2\,{\hat t}^{13}+{59\over148\,635\,648}t\,{\hat t}^{15}
    +{1\over339\,738\,624}{\hat t}^{17}+\ldots \ . }}
Here we have also inserted the closed-string mirror map~\MirrorClosedOne\ so as to eliminate the algebraic variable, $\psi$. 

Finally we need to identify the relative period encoding the superpotential, $W$. Looking again at the extracted prepotential, $\cF$, we notice that it transforms under the $\IZ_8$~transformation, $\psi\rightarrow \eta\,\psi$, as $\cF \rightarrow \eta^6 \cF$. On the other hand we know that the closed-string moduli space is equipped with a line bundle, $\cL$, whose first Chern class equals the K\"ahler form on the moduli space. Furthermore the holomorphic prepotential is a section of the line bundle, $\cL^2$ \refs{\StromingerPD,\BershadskyCX}, whereas the holomorphic superpotential, which arises from semi-periods and thus encodes the orbifold disk amplitudes, constitutes a section of the line bundle, $\cL$ \BershadskyCX. Therefore we expect the D5-brane superpotential to transform as $W\rightarrow \eta^3 W$ with respect to the mentioned $\IZ_8$~transformation.\foot{This is in contrast to the local geometries discussed in refs.~\refs{\BouchardYS,\BouchardGU}. There the superpotential is required to be invariant with respect to the monodromy group.} Hence up to an undetermined overall numerical normalization the D-brane effective superpotential in flat coordinates is given in terms of the semi-period, $\rel\Pi^5$, by
\eqn\WFlatOne{ W(t,\hat t)\,=\, 
  \left. {\rel\Pi^5(\psi,\phi)\over\rel\Pi^0(\psi)}\right|_{\psi=\psi(t),\,\phi=\phi(t,\hat t)} \ . }
It enjoys the expansion in flat coordinates
\eqn\WFlatExpOne{\eqalign{ W(t,\hat t)\,=&\,
   %% degree 3
   -4\,t\,{\hat t}+{1\over3}{\hat t}^3
   %% degree 11
   -{5\over24}t^5\,{\hat t}+{73\over576}t^4\,{\hat t}^3\cr
   &\quad
   -{29\over720}t^3\,{\hat t}^5
   +{7\over960}t^2\,{\hat t}^7-{23\over32\,256}t\,{\hat t}^9
   +{89\over3\,041\,280}{\hat t}^{11}+\ldots   \ . }}

Let us now analyze and discuss our calculated results in detail. First of all we observe that the leading two terms of the effective superpotential~\WFlatExpOne\ precisely agree with the effective cubic superpotential calculated in ref.~\KnappUW. The latter superpotential is derived by analyzing obstructions to deformations of matrix factorizations, which model the considered D5-brane geometry. However, by construction the obtained deformation superpotential is not given in terms of flat coordinates, and hence the sub-leading terms of the flat superpotential, $W(t,\hat t)$, are not visible. The critical points of the effective superpotential~\WFlatOne\ with respect to the open-string flat coordinate, $\hat t$, are determined by
\eqn\WCPoints{
  0\,=\, \partial_{\hat t}W(t,\hat t)\,=\,{\chi^5\left(\phi(\psi-8\phi)\right)\over\rel\Pi^0(\psi)}
  {\partial\phi(t,\hat t)\over\partial{\hat t}} \ , }
where we used eq.~\PeriodCondOne. Thus due to eq.~\SolChi\ the critical points of the flat superpotential with respect to the flat coordinate, $\hat t$, are located at $\phi_+(t,\hat t)=0$ and $\phi_-(t,\hat t)=8\,\psi(t)$. Hence the computed effective superpotential, $W(t,\hat t)$, reproduces the correct critical loci~\CPoints, at which the D5-brane becomes supersymmetric and wraps one of the holomorphic two cycles, $C_\pm$. We should emphasize that the agreement with the cubic deformation superpotential and the replication of the critical loci, $\phi_\pm$, are non-trivial confirmations of our computational methods.

With the critical loci, $\phi_\pm$, of the effective superpotential, $W(t,\hat t)$, at hand it is straight forward to compute the domain wall tension between the supersymmetric D5-brane wrapping the two cycle, $C_+$, and the supersymmetric D5-brane wrapping the two cycle, $C_-$. Since the only dependence of the effective superpotential, $W(t,\hat t)$, on the open-string modulus is encoded in the relative period, $\rel\Pi^5$, the relevant information about the domain wall tension is captured in the domain wall period, $\tau(\psi)$, which is the difference of the relative period, $\rel\Pi^5$, evaluated at the critical points, $\phi_\pm$. Note that, in order to extract the domain wall period, it is necessary that the critical points, $\phi_\pm$, are in the radius of convergence of the stated relative period, $\rel\Pi^5$. As the two critical points, $\phi_\pm$, approach each other at the $\IZ_8$-orbifold point, its vicinity is suitable for this calculation. Starting from the expansion~\PureRelPeriodsOne\ we arrive after a few steps of algebra at the domain wall period
\eqn\DWTensionOrbOne{
  \tau(\psi)\,=\,\rel\Pi^5(\psi,\phi_+) - \rel\Pi^5(\psi,\phi_-)\,=\,
     \sum_{k=0}^{+\infty} {2\,\Gamma(1/2+k)^4 \over \pi^{3/2}\,\Gamma(5/2+4k)} (4\psi)^{8k+3} \ . }
Alternatively, using the gamma function identities, $\Gamma(z)\,\Gamma(1-z)={\pi\over \sin(\pi z)}$ and $\Gamma(z)\,\Gamma(z+{1\over2})=\sqrt{\pi}\,2^{1-2z}\,\Gamma(2z)$, the domain wall period, $\tau(\psi)$, can also be written as
\eqn\DWTensionOrbOneRW{
  \tau(\psi)\,=\,
      {4\pi^2\over\psi}\sum_{k=1}^{\infty}
      {\Gamma(-8k+5)\over \Gamma(-4k+3)\,\Gamma(-k+3/2)^4}\,
      \left(1\over (8\psi)^8\right)^{-k+{1\over 2}} \ , }
which is a solution of the inhomogeneous Picard-Fuchs equation
\eqn\DWPFEquation{ \cL^{\rm bulk} \tau(\psi)\,=\, {3\over8\psi}\,\sqrt{(8\psi)^8} \ . }

Let us pause to compare this result with the literature. In ref.~\KnappUW\ (\cf also ref.~\KreflSJ) the domain wall tension period is computed in the vicinity of the large complex structure point of the closed string modulus, $\psi$. This result is then analytically continued to the $\IZ_8$-orbifold point, where the domain wall period splits into a contribution arising from closed string periods and into a contribution, $\tau(\psi)$, intrinsic to the domain wall tension at the orbifold locus. The latter part, $\tau(\psi)$, however, agrees precisely with the domain wall tension period~\DWTensionOrbOneRW.\foot{Compared to ref.~\KnappUW\ the expression has an additional factor of $\psi^{-1}$. This factor is traced back to fact that the normalizing period, $\rel\Pi^0(\psi)$, differs also by the same factor, $\psi^{-1}$.} The agreement can also be seen by comparing the inhomogeneous Picard-Fuchs equations. Namely by rewriting the inhomogeneous Picard-Fuchs equation~\DWPFEquation\ in terms of the large complex structure coordinate, $z=(8\psi)^{-8}$, we find the inhomogeneous Picard-Fuchs equations for domain wall tensions in the large complex structure regime stated in refs.~\KreflSJ.\foot{Compared to ref.~\KreflSJ\ we again need to take into account an additional factor of $\psi^{-1}$ arising from the normalization of the period, $\rel\Pi^0(\psi)$.} We should stress the significance of this result. From the variation of mixed Hodge structure of relative three-form periods we have obtained an effective superpotential in flat coordinates. This superpotential encodes disk instanton corrected domain wall tensions of the mirror geometry, which are computed by different means in ref.~\refs{\WalcherRS,\MorrisonBM,\KnappUW,\KreflSJ}. Thus extracting the quantum corrected domain wall period is a highly non-trivial consistency check on the flat effective superpotential, $W(t,\hat t)$, and in particular on its sub-leading terms.

%%%%%%%%%%%%%%%%%%%%%%%%%%%%%%%%%%
\tabinsert\NNExpOne{The table lists some orbifold disk invariants, $N_{k,n}^{(0,1)}$, for the analyzed D5-brane geometry in the degree eight Calabi-Yau hypersurface in ${\bf\IW\IP^{4}_{(1,1,1,1,4)}/(\IZ_8)^2\times\IZ_2}$. These invariants are normalized such that $N_{1,1}^{(0,1)}=1$, and, furthermore, are ambiguous up to the numerical normalizations of the open- and closed-string flat coordiantes, $t$ and $\hat t$.}
{\centerline{\vbox{
\offinterlineskip
\tabskip=0pt\halign{
\vrule height10pt depth6pt#\tabskip=2.5pt plus 1fil\strut
   &\hfil#\hfil&\vrule#&\hfil#\hfil&\hfil#\hfil&\hfil#\hfil&\hfil#\hfil&\hfil#\hfil&
   \hfil#\hfil&\hfil#\hfil&\hfil#\hfil&\hfil#\hfil&\tabskip=0pt\vrule#\cr
\noalign{\hrule}
&$\scriptstyle N_{k,n}^{(0,1)}$ &&
$\ \ k=1\ \ $ & $\ \quad2\quad\ $ & $\ \quad3\quad\ $ & $\ \quad4\quad\ $ & $\ \quad5\quad\ $ 
& $\ \quad6\quad\ $ & $\ \quad7\quad\ $ & $\ \quad8\quad\ $ & $\ \quad9\quad\ $ &\cr
\noalign{\hrule height1pt}
&$\,n=0$ && $0$ & $0$ & $-{1\over2}$ & $0$ & $0$ & $0$ & $0$ & $0$ & $0$&\cr
&$\phantom{\,n=\,}1$ && $1$ & $0$ & $0$ & $0$ & $0$ & $0$ & $0$ & $0$ & ${1035\over 16}$&\cr
&$\phantom{\,n=\,}2$ && $0$ & $0$ & $0$ & $0$ & $0$ & $0$ & $-{147\over16}$ & $0$ & $0$&\cr
&$\phantom{\,n=\,}3$ && $0$ & $0$ & $0$ & $0$ & ${29\over24}$ & $0$ & $0$ & $0$ & $0$&\cr
&$\phantom{\,n=\,}4$ && $0$ & $0$ & $-{73\over384}$ & $0$ & $0$ & $0$ & $0$ & $0$ & $0$&\cr
&$\phantom{\,n=\,}5$ && ${5\over96}$ & $0$ & $0$ & $0$ & $0$ & $0$ & $0$ & $0$ & ${315\,647\over512}$&\cr
&$\phantom{\,n=\,}6$ && $0$ & $0$ & $0$ & $0$ & $0$ & $0$ & $\scriptstyle-{1\,308\,259\over46\,080}$ & $0$ & $0$&\cr
&$\phantom{\,n=\,}7$ && $0$ & $0$ & $0$ & $0$ & ${91\over60}$ & $0$ & $0$ & $0$ & $0$&\cr
&$\phantom{\,n=\,}8$ && $0$ & $0$ & $\scriptstyle-{94\,379\over860\,160}$ & $0$ & $0$ & $0$ & $0$ & $0$ & $0$&\cr
\noalign{\hrule} }}}}
%%%%%%%%%%%%%%%%%%%%%%%%%%%%%%%%%%

Since we have managed to extract a uniquely distinguished set of flat relative periods in the vicinity of the $\IZ_8\times\IZ_2$-orbifold singularity we are also able to extract orbifold invariants, namely the expansions of the prepotential~\FlatFOne\ and the effective superpotential~\WFlatExpOne\ in terms of flat coordinates yield closed- and open-string orbifold invariants respectively. In particular the flat effective superpotential~\WFlatExpOne\ encodes the orbifold disk amplitudes, $N_{k,n}^{(0,1)}$, which are defined by \refs{\BouchardYS,\BriniRH,\BouchardGU}
\eqn\EquInvW{ W(t,\hat t)\,=\,\sum_{k,n} {1\over k!} N_{k,n}^{(0,1)}\,t^n\,{\hat t}^k \ . }
We have collected some of these orbifold disk invariants in \ltab\NNExpOne. Note that for the invariants in the table the effective superpotential, $W(t,\hat t)$, is rescaled such that the disk amplitude, $N_{1,1}^{(0,1)}$, is normalized to one, and furthermore the listed invariants are only defined up to the undetermined overall numerical normalization of the flat coordinates, $t$ and $\hat t$. As explained before the normalization of the closed flat coordinate, $t$, is established by explicitly extracting closed-string genus zero orbifold Gromov-Witten invariant in the mirror topological A-model, whereas the normalization of the open flat coordinate, $\hat t$, and the normalization of the superpotential, $W$, is obtained by matching the stated orbifold disk invariants with the open orbifold Gromov-Witten invariants in the topological A-model of the mirror configuration. Thus it would be interesting to pin down the normalization ambiguities and to check our results by performing an appropriate localization computation directly in the mirror topological A-model.

%%%%%%%%%%%%%%%%%%%%%%
\subsec{Large complex structure vicinity}
\subseclab\secLCS
%%%%%%%%%%%%%%%%%%%%%%
An obvious task is to analyze the open/closed Picard-Fuchs equations in the vicinity of the large complex structure point. The effective superpotential in this regime potentially encodes disk instantons of the mirror D6-brane configuration. Instanton generated superpotentials appear in the mirror type~IIA theory for chiral multiplets of the open-string sector that are massless and give rise to flat directions at the large volume point \refs{\WittenFB,\KachruIH}. For the example at hand we do not expect any open/closed disk instantons associated to the (obstructed) open-string modulus, $\phi$, because it interpolates between the two critical points, $\phi_\pm$, separated by a domain wall. The domain wall tension, however, remains finite at the large complex structure point \refs{\KnappUW,\KreflSJ}, and therefore the obstructed modulus, $\phi$, does not give rise to a flat direction, which would  indicate the appearance of disk instantons. 

Nevertheless let us briefly discuss the singularity structure of the connection matrices~\MPhiOne\ and \MPsiOne\ of the linear Gauss-Manin system in the vicinity of the large complex structure point. First we observe that the large complex structure locus, $\psi=\infty$, intersects the discriminant locus, $D_2=0$, at the point, $(\psi,\phi)=(\infty,\infty)$, in the open/closed moduli space. By analyzing the connection matrices in the vicinity of this intersection point we observe that the locus, $\phi=\infty$, is singular for any value of $\psi$. Thus actually three boundary divisors of the moduli space meet at the point, $(\psi,\phi)=(\infty,\infty)$. This intricate singularity structure becomes also apparent by looking at the degeneration of the D5-brane divisor, $V$, itself. The defining polynomial of the Calabi-Yau degenerates at the large comlex structure point to to the monomial, $P(\infty)=x_1 x_2 x_3 x_4 x_5$. On the other hand in the limit, $|\phi|\rightarrow+\infty$, the D5-brane divisor turns into $Q(\infty)=x_1 x_2 x_3 x_4$. Hence the intersection locus of the two polynomials, $P$ and $Q$, which is complex two-dimensional at a generic point in the moduli space, obtains at the point, $(\psi,\phi)=(\infty,\infty)$ complex three-dimensional components, which are given by the three-dimensional weighted projective spaces, $\IW\IP^3_{(1,1,1,4)}$, embedded into the ambient space, $\IW\IP^4_{(1,1,1,1,4)}$.

A detailed discussion of the large complex structure regime presumably requires to resolve the intersection point of the three boundary divisors in the open/closed moduli space along the lines ref.~\refs{\CandelasDM,\CandelasHW}. This analysis, however, is beyond the scope of this work, but we hope to come back to this issue elsewhere.

%%%%%%%%%%%%%%%%%%%%%%
\newsec{D5-branes in the mirror quintic threefold}
%%%%%%%%%%%%%%%%%%%%%%
Now we investigate a certain class of D5-branes in the mirror quintic threefold in the projective space, $\IC\IP^4$. Analogously to the example in the previous section we derive the open/closed Picard-Fuchs equations, and we compute the effective superpotential in flat coordinates. From the effective superpotential we also extract a domain wall tension, which agrees with the result obtained in ref.~\refs{\WalcherRS,\MorrisonBM}. In the analysis we proceed analogously to the previous example. Therefore we mainly emphasis the differences in this section, defer the calculations to Appendix~B, and refer for further explanation to Section~\secExampleOne.

%%%%%%%%%%%%%%%%%%%%%%
\subsec{The Calabi-Yau and D-brane geometry}
%%%%%%%%%%%%%%%%%%%%%%
In this section the bulk geometry of interest is given by the mirror of the quintic threefolds, $Y$, in the projective space, $\IC\IP^4$. The family of quintics depends on $101$ complex structure moduli and one K\"ahler modulus, whence the family of mirror quintics, $Y$, have one complex structure and $101$ K\"ahler moduli, as explained in detail in ref.~\CandelasRM. It is defined by the homogeneous degree five polynomial,
\eqn\PTwo{ P(\psi)\,=\,x_1^5+x_2^5+x_3^5+x_4^5+x_5^5-5\,\psi\,x_1x_2x_3x_4x_5 \ , }
in the $\IZ_5^3$~orbifold of the projective space, $\IC\IP^4$ \refs{\GreeneUD,\CandelasRM}, and the algebraic modulus, $\psi$, parametrizes the one-dimensional complex structure moduli space. Furthermore the $\IZ_5^3$~orbifold is generated by
\eqn\OrbGenTwo{g_1=(1,0,0,0,4) \ , \quad g_2=(0,1,0,0,4) \ , \quad g_3=(0,0,1,0,4) \ , \quad g_4=(0,0,0,1,4) \ ,}
where, for instance, the generator, $g_1$, acts on the homogeneous projective coordinates as\foot{Since the transformation, $g_1g_2g_3g_4$, induces a homogeneous rescaling of the projective coordinates there are really  only three independent generators~\OrbGenTwo.}
\eqn\GenActionTwo{
  g_1: \ [\,x_1 : x_2 : x_3 : x_4 : x_5\,] \mapsto [\,\rho\,x_1 : x_2 : x_3 : x_4 : \rho^4x_5\,] \ ,
  \quad \rho\equiv e^{2\pi i/5} \ . }
Resolving the resulting orbifold singularities gives rise to the smooth family of mirror quintics depending on $101$~K\"ahler moduli.

The divisor, $V$, for the D5-brane geometry is specified by the homogeneous degree-four polynomial, $Q$. As there are only two monomials of degree four with definite $(\IZ_5)^3$~charges we arrive at
\eqn\PolDivTwo{ Q(\phi)\,=\,x_5^4-\phi\,x_1 x_2 x_3 x_4 \ , }
where the parameter, $\phi$, is the algebraic open-string modulus.

The geometry specified by the homogeneous polynomials, $P(\psi)$ and $Q(\phi)$, is invariant with respect to the discrete $\IZ_5$~symmetry, which acts on the open/closed algebraic moduli as
\eqn\DiscreteFive{  \pmatrix{ \psi \cr \phi} \mapsto \rho \pmatrix{ \psi \cr \phi } \ , \quad \rho\equiv e^{2\pi i/5} \ .}
The projective coordinate transformation, $x_1\mapsto\rho^{-1}x_1$, compensates the generator~\DiscreteFive\ and hence establishes this $\IZ_5$~symmetry on the level of the algebraic variables, $\psi$ and $\phi$. As a consequence the open/closed string moduli space arises really as the $\IZ_5$~orbifold of the covering space parametrized by the variables, $\psi$ and $\phi$.

The next step is to relate the divisor, $Q$, to the two supersymmetric D5-branes  appearing in refs.~\refs{\WalcherRS,\MorrisonBM}. The geometric embedding, $C_\pm$, into the Calabi-Yau threefold, $Y$, are specified by \MorrisonBM
\eqn\PolCtwo{ C_\pm\,=\, \left\{ x_1+x_2=0\, , x_3+x_4=0\, , x_5^2\pm\sqrt{5\psi}\,x_1x_3=0\right\} \subset Y \ , }
together with their images under the $(\IZ_5)^3$-orbifold group. By inserting the first two conditions, $x_1+x_2=0$ and $x_3+x_4$, into the divisior, $Q(\phi)$, we observe that the holomorphic two cycles, $C_\pm$, are simultaneously contained in the divisor, $Q(\phi)$, at the critical point,
\eqn\CPoint{ \phi_0\,=\,5\,\psi \ . }
As a consequence we expect that the divisor, $V$, specified by the polynomial, $Q(\phi)$, describes a configuration of D5-branes, which at the critical locus, $\phi_0$, wraps both holomorphic two cycles, $C_\pm$. The fact that the polynomial, $Q(\phi_0)$, does not discriminate between the cycles, $C_+$ and $C_-$, gives us less control over the open-string moduli dependence of the described D5-brane configuration compared to the examples studied in Section~\secExampleOne. In particular by moving away from the critical locus the open-string modulus, $\phi$, interlocks the deformations of both D5-brane cycles. Despite these subtleties we are still able to extract the correct domain wall tension by evaluating the derived effective superpotential at the critical points, $(\phi^{1/2})_\pm=\pm\sqrt{5\psi}$.

%%%%%%%%%%%%%%%%%%%%%%
\subsec{Solutions to the open/closed Picard-Fuchs equations}
%%%%%%%%%%%%%%%%%%%%%%
In this section we derive the open/closed Picard-Fuchs system of partial differential equations and solve them in the vicinity of the $\IZ_5$-orbifold locus, \ie in the vicinity of $\psi=0$ and $\phi=0$. We perform our analysis along the lines of Section~\secExampleOne.

Starting from the relative holomorphic three form, $\rel\Omega(\psi,\phi)$, we construct by Griffiths transversality a basis for the relative cohomology group, $H^3(Y,V)$. This basis is given by
\eqn\BasisTwo{
  \rel\pi \equiv
  \left( \pi_{3,3} , \pi_{2,3} , \pi_{1,3} , \pi_{0,3} , \pi_{2,4} , \pi_{1,4} , \pi_{0,4} \right) =
  \left( \rel\Omega , \partial_\psi\rel\Omega , \partial_\psi^2\rel\Omega , \partial_\psi^3\rel\Omega ,
          \partial_\phi\rel\Omega , \partial_\psi\partial_\phi\rel\Omega ,
          \partial_\psi^2\partial_\phi\rel\Omega \right) \ , }
where the indices are again labelled in accord with the variation of mixed Hodge structrure depicted in the diagram~\VarMixed. We explicitly represent the basis elements~\BasisTwo\ in terms of relative three-form residue integrals, which allow us to calculate the associated linear Gauss-Manin system
\eqn\GMSysTwo{
  \left(\partial_\psi - M_\psi\right) \rel\pi\,\simeq\,0 \ , \quad
  \left(\partial_\phi - M_\phi\right) \rel\pi\,\simeq\,0 \ . }
The derivation of the connection matrices, $M_\psi$ and $M_\phi$, is deferred to Appendix~B. The result of this tedious analysis yields
\eqn\MPsiTwo{
  M_\psi\,=\,\pmatrix{
                         0 & 1 & 0 & 0 & 0 & 0 & 0 \cr
                         0 & 0 & 1 & 0 & 0 & 0 & 0 \cr
                         0 & 0 & 0 & 1 & 0 & 0 & 0 \cr
                         {\psi\over D_1} & {15\psi^{2}\over D_1} & {25\psi^{3}\over D_1} & {10\psi^{4}\over D_1} & {-\phi T_{1}\over 16 D_1D_2} & {-\phi T_{2}\over 16 D_1D_2} & {-\phi T_{3}\over 16 D_1D_2} \cr
                         0 & 0 & 0 & 0 & 0 & 1 & 0 \cr
                         0 & 0 & 0 & 0 & 0 & 0 & 1 \cr
                         0 & 0 & 0 & 0 & {125\,\phi(\phi-5\psi)\over D_2} & {-175\phi(\phi-5\psi)^{2}\over D_2} & {30\,\phi(\phi-5\psi)^{3}\over  D_2}  } \ , }
and
\eqn\MPhiTwo{
  M_\phi\,=\, \pmatrix{
                         0 & 0 & 0 & 0 & 1 & 0 & 0 \cr
                         0 & 0 & 0 & 0 & 0 & 1 & 0 \cr
                         0 & 0 & 0 & 0 & 0 & 0 & 1 \cr
                         0 & 0 & 0 & 0 & {125\,\phi(\phi-5\psi)\over D_2} & {-175\phi(\phi-5\psi)^{2}\over D_2} & {30\,\phi(\phi-5\psi)^{3}\over D_2} \cr
                         0 & 0 & 0 & 0 & -{3\over 4\phi} & -{\phi-\psi\over 4\phi} & 0 \cr
                         0 & 0 & 0 & 0 & 0 & -{1\over 2\phi} & -{\phi-\psi\over 4\phi} \cr
                         0 & 0 & 0 & 0 & -{125(\phi-\psi)(\phi-5\psi)\over 4 D_2} & {175(\phi-\psi)(\phi-5\psi)^{2}\over 4 D_2} & -{1\over 4\phi}-{15(\phi-\psi)(\phi-5\psi)^{3}\over 2 D_2} } \ , }
in terms of the polynomials
\eqn\threeT{\eqalign{T_{1}=\,&\phi  \left(8000-\phi  (\phi -5 \psi ) \psi  \left(61 \phi ^2-790 \psi  \phi
   +2825 \psi ^2\right)\right)-16384\, \psi \ ,\cr
   T_{2}=\,&57375 \phi ^2 \psi ^5-34000 \phi ^3 \psi ^4+7190 \phi ^4 \psi ^3-8 \left(79
   \phi ^5+14336\right) \psi ^2 \cr
   &+\phi  \left(19 \phi ^5+95936\right) \psi
   -11200\, \phi ^2\ ,\cr
   T_{3}=\,&22625 \phi ^2 \psi ^6-16325 \phi ^3 \psi ^5+4490 \phi ^4 \psi ^4-2 \left(293
   \phi ^5+49152\right) \psi ^3 \cr
   & +\phi  \left(37 \phi ^5+112768\right) \psi
   ^2  -\phi ^2 \left(\phi ^5+26624\right) \psi +1920\, \phi ^3\ ,}}
and the discriminants
\eqn\DiscQuintic{D_1\,=\,1-\psi^5 \ , \quad D_2\,=\, \phi (\phi-5 \psi)^4-256 \ . }
The discriminant locus, $D_1=0$, corresponds to the familiar conifold point of the bulk geometry, whereas the locus, $D_2=0$, describes again a singularity in the open sector. Namely the intersection locus of the two polynomials, $P$ and $Q$, fails to be transversal at some points in the ambient projective space, $\IC\IP^4$.

It is straight forward to check that the Gauss-Manin connection, $\nabla_\phi\equiv\partial_\phi-M_\phi$ and $\nabla_\psi\equiv\partial_\psi-M_\psi$, is integrable, \ie the Gauss-Manin connection is flat, $[\nabla_\phi,\nabla_\psi]\equiv 0$. Integrability ensures that the associated open/closed Picard-Fuchs system of differential equations~\PFPeriod\ for the relative periods is solvable. From the Gauss-Manin system~\GMSysTwo\ we extract three partial differential Picard-Fuchs operators of the form
\eqn\PFTwo{
   \cL_1\,=\, \widetilde\cL_1\, \partial_\phi \ , \quad
   \cL_2\,=\, \widetilde\cL_2\, \partial_\phi \ , \quad
   \cL_3\,=\, \cL^{\rm bulk} + \cL^{\rm bdry} \ .}
Here the operators, $\widetilde\cL_1$ and $\widetilde\cL_2$, read
\eqn\PFSubTwo{\eqalign{
   \widetilde\cL_1 \,&=\, (\psi-\phi)\,\theta_{\psi}-4\psi\,\theta_{\phi}-3\,\psi \ , \cr
   \widetilde\cL_2 \,&=\,\partial^{2}_{\psi}\partial_{\phi}+\left({1\over 4\phi}+{15\over 2D_{2}}\,(\phi-\psi)(\phi-5\psi)^{3}\right)\partial^{2}_{\psi}-{175\over 4D_{2}}\,(\phi-\psi)(\phi-5\psi)^{2}\, \partial_{\psi}\cr
   &\qquad +{125\over 4D_{2}}\,(\phi-\psi)(\phi-5\psi)\ , }}
and the operators, $\cL^{\rm bulk}$ and  $\cL^{\rm bdry}$, are
\eqn\PFWholeTwo{\eqalign{
   \cL^{\rm bulk} \,&=\,\theta_{\psi}(\theta_{\psi}-1)(\theta_{\psi}-2)(\theta_{\psi}-3)
   -\psi^{5}(\theta_{\psi}+1)^{4} \ , \cr
   \cL^{\rm bdry} \,&=\,\psi^4\,D_1\,\partial^{4}_{\psi}+{\psi^4\phi\over16D_2}\Big(T_{3}\,\partial^{2}_{\psi}\partial_{\phi}+T_{2}\,\partial_{\psi}\partial_{\phi}+T_{1}\,\partial_{\phi}\Big) \ . }}

The solutions to the bulk Picard-Fuchs operator, $\cL^{\rm bulk}$, in the vicinity of the orbifold point, $\psi=0$, are determined by the hypergeometric functions
\eqn\BulkSolTwo{\eqalign{
  \rel\Pi^0(\psi)\,&=\,\HGF{4}{3}\left({1\over5},{1\over5},{1\over5},{1\over5};{2\over5},{3\over5},{4\over5};\psi^{5}
  \right) \ , \cr
  \rel\Pi^1(\psi)\,&=\,\psi\,\HGF{4}{3}\left({2\over5},{2\over5},{2\over5},{2\over5};{3\over5},{4\over5},{6\over5};
  \psi^{5}\right) \ , \cr
  \rel\Pi^2(\psi)\,&=\,{\psi^{2}\over 2!}\,\HGF{4}{3}\left({3\over5},{3\over5},{3\over5},{3\over5};{4\over5},{6\over5},{7\over5}
  ;\psi^{5}\right) \ , \cr
  \rel\Pi^3(\psi)\,&=\,-{\psi^{3}\over 3!}\,\HGF{4}{3}\left({4\over5},{4\over5},{4\over5},{4\over5};{6\over5},{7\over5},{8\over5};\psi^{5}
  \right) \ , }}
or in terms of the power series
\eqn\BulkSolTwoExp{
  \rel\Pi^\alpha(\psi)\,=\,{c_{\alpha}\over \alpha!\,5^{\alpha}\,\Gamma\big({\alpha+1\over 5}\big)^{5}}\,\sum_{k=0}^{\infty}\,{\Gamma\big(k+{\alpha+1\over5}\big)^{5}\over \Gamma(5k+\alpha+1)}\,(5\psi)^{5k+\alpha}\ , \qquad \alpha\,=\,0,1,2,3 \ , }
with radius of convergence $|\psi|<1$. As in the previous example all the constants, $c_\alpha$, are one except for $c_3=-1$. 

In order to calculate the relative periods resulting from semi-periods we proceed analogously to Section~\secExampleOne. That is to say we first solve the variational sub-system~\VarSubSys\ governed by the two differential operators, $\widetilde\cL_1$ and $\widetilde\cL_2$. The operator, $\widetilde\cL_1$, constraints a solution of the sub-system to have the form
\eqn\SolLOneQuin{\chi(\psi,\phi)\,\equiv\,\phi^{-{1\over4}}(5\psi-\phi)^2\,\lambda(u) \quad {\rm with}
   \quad u\,\equiv\,\phi(5\psi-\phi)^4 \ , }
whereas the second operator, $\widetilde\cL_2$, applied to this ansatz yields for the function, $\lambda(u)$, the ordinary differential equation
\eqn\SolLTTwoQuin{
  \left[\theta_u \Big(\theta_{u}+{1\over2}\Big)\Big(\theta_{u}+{1\over4}\Big)-{u\over256}\Big(\theta_{u}+{3\over4}\Big)^{3} \right]\,{\lambda}(u) \,=\,0 \ . }
A complete set of solutions to this differential equation of hypergeometric type reads
\eqn\Sollambda{\eqalign{
  \lambda^4(u)\,&=\, u^{-1/2}\,\HGF{3}{2}\left({1\over4},{1\over4},{1\over4};{2\over4},{3\over4};{u\over256}\right)  \ , \cr
  \lambda^5(u)\,&=\, u^{-1/4}\,\HGF{3}{2}\left({1\over2},{1\over2},{1\over2};{3\over4},{5\over4};{u\over256}\right) \ , \cr
  \lambda^6(u)\,&=\, \HGF{3}{2}\left({3\over4},{3\over4},{3\over4};{5\over4},{6\over4};{u\over256}\right) \  . }}
The power series of these solutions are convergent for $|u|<256$ and are given by
\eqn\SollambdaExpand{
  \lambda^{4+\ell}(u)\,=\,{1\over \Gamma\big({1+\ell\over 4}\big)^{4}}\,\sum_{k=0}^{\infty}\,{\Gamma^{4}\big(k+{1+\ell\over 4}\big)\over \Gamma(4k+\ell+1)}\, u^{k+{\ell-2\over4}}\ ,\qquad \ell\,=\,0,1,2 \ . }

Due to the integrability of the open/closed Picard-Fuchs system~\PFTwo\ we are able to derive the three relative periods, $\rel\Pi^{\hat\alpha}$, by integrating the relations, $\partial_\phi\rel\Pi^{\hat\alpha}=\chi^{\hat\alpha}(\psi,\phi)=\phi^{-{1\over 4}}(5\psi-\phi)^2\,\lambda^{\hat\alpha}\left(\phi(5\psi-\phi)^4\right)$. In this integration process we must not forget to take into account the possibility of non-trivial integration constants, $C^{\hat\alpha}(\psi)$. A detailed analysis, however, reveals that the periods
\eqn\PureRelPeriodsIntTwo{
  \rel\Pi^{\hat\alpha}(\psi,\phi)\,=\,\int_0^{\phi}
   \phi^{-{1\over 4}}(5\psi-\zeta)^2\,\lambda^{\hat\alpha}\left(\zeta(5\psi-\zeta)^4\right)\,d\zeta \ , }
furnish indeed the three linearly independent solutions to the whole open/closed Picard-Fuchs system~\PFTwo. The power series of these solutions in the vicinity of the orbifold locus become
\eqn\PureRelPeriodsExpTwo{\eqalign{
  \rel\Pi^4(\psi,\phi)\,&=\,{4\over \Gamma^{4}(1/4)}\sum_{k=0}^{\infty}\sum_{n=k}^{5k}{(-1)^{n-k}\,\Gamma^{4}(k+1/4)\over (4n+1)\,(5k-n)!\,(n-k)!}\,\phi^{n+1/4}\,(5\psi)^{5k-n}\ , \cr
  \rel\Pi^5(\psi,\phi)\,&=\,{2\over \pi^{2}}\sum_{k=0}^{\infty}\sum_{n=k}^{5k+1}{(-1)^{n-k+1}\,\Gamma^{4}(k+1/2)\over (2n+1)\,(5k-n+1)!\,(n-k)!}\,\phi^{n+1/2}\,(5\psi)^{5k-n+1} \ , \cr
  \rel\Pi^6(\psi,\phi)\,&=\,{8\over \Gamma^{4}(3/4)}\sum_{k=0}^{\infty}\sum_{n=k}^{5k+2}{(-1)^{n-k}\,\Gamma^{4}(k+3/4)\over (4n+3)\,(5k-n+2)!\,(n-k)!}\,\phi^{n+3/4}\,(5\psi)^{5k-n+2}\ . }}
These expansions are convergent for $|\phi(5\psi-\phi)^4|<256$. 

%%%%%%%%%%%%%%%%%%%%%%
\subsec{Effective superpotential and domain wall tension}
%%%%%%%%%%%%%%%%%%%%%%
With the relative three-form periods~\BulkSolTwo\ and \PureRelPeriodsExpTwo\ at hand we are now ready to extract distinguished flat open and closed coordinates so as to formulate the flat effective superpotential. The analysis is performed in the vicinity of the $\IZ_5$-orbifold point and therefore parallels our investigations in Section~\secExampleOne. Hence for additional explanations we again refer the reader to the previous section.

So as to determine the flat coordinates, the prepotential and the effective superpotential, we apply the same criteria discussed thoroughly in the context of the previous example. Namely we require that the flat periods must not be singular at the orbifold point, $\psi=\phi=0$, and furthermore the flat periods should furnish one-dimensional representations with respect to the $\IZ_5$-monodromy group at the orbifold locus. These requirements yield, up to numerical constants, the open/closed flat coordinates
\eqn\FlatOCTwo{
  t(\psi)\,=\,{\rel\Pi^1(\psi)\over\rel\Pi^0(\psi)} \ , \quad
  \hat t(\psi,\phi)\,=\,{\rel\Pi^4(\psi,\phi)\over\rel\Pi^0(\psi)} \ , }
the prepotential, $\cF(t)$, and the effective superpotential, $W(t,\hat t)$,
\eqn\FlatFWTwo{
  \cF(t)\,=\,{1\over2}\left.{\rel\Pi^3(\psi)\over\rel\Pi^0(\psi)}\right|_{\psi=\psi(t)}
     +\,{t\over2}\left.{\rel\Pi^2(\psi)\over\rel\Pi^0(\psi)}\right|_{\psi=\psi(t)} \ , \quad
 W(t,\hat t)\,=\,
  \left. {\rel\Pi^5(\psi,\phi)\over\rel\Pi^0(\psi)}\right|_{\psi=\psi(t),\,\phi=\phi(t,\hat t)} \ . }
in terms of the open/closed mirror maps. The first few terms in the expansion of the closed string mirror map read
\eqn\MirrorClosedTwo{
  \psi(t)\,=\,t-{13\over360}\,t^{6}-{31\,991\over9\,979\,200}\,t^{11}
    -{294\,146\,129\over326\,918\,592\,000}\,t^{16}+\cdots \ . }
For the leading terms of the open mirror map we arrive at
\eqn\MirrorOpenTwo{\eqalign{
  \phi^{1\over 4}(t,\hat t)\,=&\,{1\over4}\,\hat{t}+{1\over480}\,t^{5}\hat{t}
   -{125\over2^{21}\cdot 3}\,t^{4}\hat{t}^{5}+{125\over2^{27}\cdot 27}\,
  t^{3}\hat{t}^{9}-{25\over2^{36}\cdot 13}\,t^{2}\hat{t}^{13}\cr
  &+{5\over2^{43}\cdot 51}\,t\,\hat{t}^{17}-{1\over2^{53}\cdot 63}\,
  \hat{t}^{21}+\cdots \ . }}
These expansions are obtained by inverting the flat coordinates~\FlatOCTwo.

%%%%%%%%%%%%%%%%%%%%%%%%%%%%%%%%%%
\tabinsert\NNExpTwo{The table lists some orbifold disk invariants, $N_{k,n}^{(0,1)}$, for the analyzed D5-brane geometry in the mirror quintic, which are extracted form the superpotential, $\widetilde W(t,\tilde t)$. These invariants are normalized such that $N_{1,1}^{(0,1)}=1$, and are ambiguous up to the normalizations of the open- and closed-string coordinates, $t$ and $\tilde t$.}
{\centerline{\vbox{
\offinterlineskip
\tabskip=0pt\halign{
\vrule height10pt depth6pt#\tabskip=2.5pt plus 1fil\strut
   &\hfil#\hfil&\vrule#&\hfil#\hfil&\hfil#\hfil&\hfil#\hfil&\hfil#\hfil&\hfil#\hfil&
   \hfil#\hfil&\hfil#\hfil&\hfil#\hfil&\hfil#\hfil&\tabskip=0pt\vrule#\cr
\noalign{\hrule}
&$\scriptstyle N_{k,n}^{(0,1)}$ &&
$\ \ k=1\ \ $ & $\ \quad2\quad\ $ & $\ \quad3\quad\ $ & $\ \quad4\quad\ $ & $\ \quad5\quad\ $
& $\ \quad6\quad\ $ & $\ \quad7\quad\ $ & $\ \quad8\quad\ $ & $\ \quad9\quad\ $ &\cr
\noalign{\hrule height1pt}
&$\,n=0$ && $0$ & $0$ & $-{1\over640}$ & $0$ & $0$ & $0$ & $0$ & $0$ & $0$&\cr
&$\phantom{\,n=\,}1$ && $1$ & $0$ & $0$ & $0$ & $0$ & $0$ & $0$ & $0$ & $0$&\cr
&$\phantom{\,n=\,}2$ && $0$ & $0$ & $0$ & $0$ & $0$ & $0$ & $0$ & $0$ & ${564\,375\over2^{34}\cdot221}$&\cr
&$\phantom{\,n=\,}3$ && $0$ & $0$ & $0$ & $0$ & $0$ & $0$ & ${70\,625\over2^{29}\cdot 39}$ & $0$ & $0$&\cr
&$\phantom{\,n=\,}4$ && $0$ & $0$ & $0$ & $0$ & $-{5375\over2^{13}\cdot 9}$ & $0$ & $0$ & $0$ & $0$&\cr
&$\phantom{\,n=\,}5$ && $0$ & $0$ & ${999\over2^{17}\cdot 5}$ & $0$ & $0$ & $0$ & $0$ & $0$ & $0$&\cr
&$\phantom{\,n=\,}6$ && $-{1\over36}$ & $0$ & $0$ & $0$ & $0$ & $0$ & $0$ & $0$ & $0$&\cr
&$\phantom{\,n=\,}7$ && $0$ & $0$ & $0$ & $0$ & $0$ & $0$ & $0$ & $0$ & ${5^5\cdot 7\,011\,679\over2^{42}\cdot1989}$&\cr
&$\phantom{\,n=\,}8$ && $0$ & $0$ & $0$ & $0$ & $0$ & $0$ & $-{5^7\cdot 55\,819\over2^{40}\cdot 273}$ & $0$ & $0$&\cr
\noalign{\hrule} }}}}
%%%%%%%%%%%%%%%%%%%%%%%%%%%%%%%%%%

Finally the holomorphic prepotential, $\cF(t)$, expressed in flat coordinates enjoys the expansion
\eqn\FFlatTwo{
  \cF(t)\,=\, {1\over 6}t^3+{1\over1008}t^8+{1195\over10378368} t^{13}
  +{6904357\over266765571072} t^{18}+{43753160719\over5523935200616448} t^{23}+\ldots \ ,}
whereas the effective superpotential, $W(t,\hat t)$, yields
\eqn\WFlatTwo{\eqalign{
  W(t,\hat t)\,=&\,-{5\over8}t\,\hat{t}^{2}+{1\over6144}\,\hat{t}^{6}+{5\over288}\,t^{6}\hat{t}^{2}-{333\over2097152}\,
  t^{5}\hat{t}^{6}+{5375\over14495514624}\,t^{4}\hat{t}^{10}\cr
  &-{70625\over168843754340352}\,t^{3}\hat{t}^{14}+{26875\over104972574127030272}\,t^{2}\hat{t}^{18}\cr
  &-{8725\over106113814420103626752}\,t\,\hat{t}^{22}+{103\over9442427122730076733440}\,\hat{t}^{26}+\cdots\ . }}

First we observe that only the square, $\hat t^2$, of the open flat coordinate, $\hat t$, enters the flat superpotential. This indicates that we really capture the product of deformations associated to the two individual branes, which at the critical locus wrap the D5-brane cycles, $C_\pm$, in eq.~\PolCtwo\ respectively. In order to compare with a single D5-brane component we introduce the superpotential, $\widetilde W$, which is given by
\eqn\WFlatAltTwo{  \widetilde W(t,\tilde t)\,=\, W(t,\hat t^2) \ . }
The leading terms of the superpotential, $\widetilde W$, coincide with the deformation superpotential computed in ref.~\HoriJA. Furthermore the modified superpotential yield the expected critical points, $\phi^{1/2}_\pm(t,\tilde t)=\pm \sqrt{5\psi(t)}$, with respect to the open-string coordinate, $\tilde t$.

Analogously to the previous example we extract the domain wall tension period, $\tau(\psi)$, in the vicinity of the orbifold locus by evaluating the superpotential period, $\rel\Pi^5$, at the critical points
\eqn\OrbPeriodTwo{\eqalign{
  \tau(\psi)\,=&\, \left.\rel\Pi^5(\psi,\phi)\right|_{\phi^{1/2}=+\sqrt{5\psi}}
    - \left.\rel\Pi^5(\psi,\phi)\right|_{\phi^{1/2}=-\sqrt{5\psi}} \cr
    \,=&\,-{2\over 5\psi \pi^{2}}\sum_{k=0}^{\infty}{\Gamma(k+1/2)^{5}\over \Gamma(5k+5/2)}\,\big[(5\psi)^{5}\big]^{k+1/2} \ . }}
Using the identity, $\Gamma(z)\,\Gamma(1-z)={\pi\over \sin(\pi z)}$, we arrive at the expression
\eqn\OrbPeriodTwoExp{
  \tau(\psi)\,=\,-{2\pi^{2}\over 5\psi}\,
  \sum_{k=0}^{\infty}{\Gamma(-5k-3/2)\over \Gamma(-k+1/2)^{5}}\, \left({1\over(5\psi)^5}\right)^{-k-1/2} \ , }
which is a solution of the inhomogeneous Picard-Fuchs equation
\eqn\DWInHomTwo{
  \cL^{\rm bulk}\tau(\psi)=-{3\pi^4\over 10\psi}\,\sqrt{(5\psi)^5} \ . }
Thus we find again agreement with the domain wall period computed in refs.~\WalcherRS, which can be seen by either comparing the inhomogeneous Picard-Fuchs equations or by directly matching the domain wall periods at the orbifold point. 

Finally we have collected the orbifold disk invariants of the superpotential, $\widetilde W(t,\tilde t)$, in \ltab\NNExpTwo. As before these invariants are defined up to numerical normalizations of the flat coordinates and the effective superpotential. Since these invariants are extracted from the superpotential, $\widetilde W(t,\tilde t)$, in terms of the open-string variable, $\tilde t$, it is tempting to identify them with the obstruction disk invariants associated to a single D5-brane components. However, in order to substantiate this claim a better understanding of the relationship of the D5-brane divisor, $V$, to its individual D5-brane components is necessary.

We should also remark that the open/closed Picard-Fuchs equations in the vicinity of the large complex structure point exhibit similar features as the open/closed Picard-Fuchs equations of the previous example as discussed in Section~\secLCS.

%%%%%%%%%%%%%%%%%%%%%%
\newsec{Conclusions}
%%%%%%%%%%%%%%%%%%%%%%
In this paper we have provided new techniques to calculate effective superpotentials for the $N=1$ low energy effective theory of type~IIB Calabi-Yau compactifications with D5-branes and fluxes. These superpotentials depend on both open- and closed-string chiral fields associated to D5-brane deformations and complex structure deformations of the internal Calabi-Yau manifold respectively. For supersymmetric configurations, \ie at the critical points of the superpotentials, these neutral chiral fields become massless and correspond to obstructed moduli fields. Thus geometrically the effective superpotentials capture the obstructions in the open-/closed-string moduli space. 

Analogously to D5-branes in local Calabi-Yau spaces discussed in refs.~\refs{\LercheCK,\LercheYW}, we have expressed the superpotentials in terms of relative periods of D5-brane boundary divisors in compact Calabi-Yau threefolds. We have demonstrated that in the context of compact Calabi-Yau spaces these relative periods are attainable from a particular type of residue integrals. As for non-compact geometries these relative periods are governed by the underlying $N=1$ special geometry \MayrXK. This structure allowed us to parametrize the open/closed moduli space in terms of flat coordinates, which we also used to express the effective superpotential. 

The effective superpotentials in flat coordinates are interesting for several reasons. First of all we have analyzed the effective superpotential in the framework of the topological B-model, and therefore the flat superpotential becomes the disk partition function of the topological A-model of the mirror D6-brane configuration in the mirror Calabi-Yau geometry  \refs{\OoguriBV,\LabastidaZP}. Thus at special points in the open/closed moduli space, namely at points where a distinguished set of flat coordinates can be determined, the flat superpotential encodes enumerative disk invariants of the A-model quantum geometry. So far most computational techniques, which are used to extract disk invariants, are mainly limited to D-branes in local Calabi-Yau configurations, whereas our approach is in particular suitable for D-branes in compact geometries.

On the other hand our methods are potentially useful in the context of string phenomenology. The interplay of the bulk geometry with D-branes and background fluxes is a crucial ingredient in constructing phenomenological viable models. Therefore by providing a handle on the effective superpotential beyond the qualitative level we possibly get new insights into the vacuum structure of type~II string compactifications with branes and background fluxes. Moreover the ability to reliably compute non-perturbative D-brane superpotentials might also shed light on aspects of dynamical supersymmetry breaking.

In this work we have also applied our methods to two examples explicitly. Our first example constituted a certain D5-brane configuration embedded in the mirror of the degree-eight hypersurface in the weighted projective space, $\IW\IP^4_{(1,1,1,1,4)}$. This setup provides for one closed-string complex structure modulus and one open-string brane modulus, and it is directly related to the geometries discussed in refs.~\refs{\KnappUW,\KreflSJ}. We derived the associated open/closed Picard-Fuchs partial differential equations and showed their integrability. Then we solved this system of differential equations in the vicinity of the orbifold point of the open/closed string moduli space. From the solutions we extracted a uniquely distinguished set of flat open-/closed-string coordinates together with their effective superpotential. As for local Calabi-Yau geometries in refs.~\refs{\BouchardYS,\BriniRH,\BouchardGU}, we determined (up to overall normalizations) from this superpotential a tower of disk orbifold invariants of the corresponding mirror geometry in the topological A-model. 

The resulting superpotential reproduces the correct critical locus in agreement with the leading order behavior of the superpotential computed by matrix-factorization techniques in ref.~\KnappUW. With our methods, however, were able to obtain the subleading corrections encoded in the flat coordinates. Finally by evaluating the superpotential at its two critical points we have calculated in the vicinity of the orbifold point the domain wall tension between the two supersymmetric D-brane configuration, which remarkably matches with the results calculated by different means in refs.~\refs{\KnappUW,\KreflSJ}. Namely, there the domain wall tension emerges as a solution of an inhomogeneous Picard-Fuchs equation. The inhomogeneous term comes from a three-chain integral, which needs to be computed analytically, whereas in our approach the domain wall tension is determined purely algebraically.

As our second example we have examined a particular family of D5-branes in the mirror quintic threefold. This setup depends on one complex structure and one D5-brane modulus and is very similar to the first example. We again computed the effective superpotential in flat coordiantes in the vicinity of the orbifold point, we extracted the corresponding orbifold disk invariants for the corresponding mirror configuration, and we determined the domain wall tension between two distinct supersymmetric D5-brane configurations in agreement with the results of refs.~\refs{\WalcherRS,\MorrisonBM}. 

There are many open questions, which we have not addressed in this work. First of all the presented derivations of the open/closed Picard-Fuchs equations are rather tedious. However, the (quasi-)homogeneity of the defining polynomials of the Calabi-Yau space and of the D-brane divisor suggests that toric techniques might help to obtain the open/closed Picard-Fuchs equations more economically.

For the two presented examples the open-string moduli are obstructed in the vicinity of the large complex structure locus. However, for configurations with open-string moduli that become unobstructed at the large complex structure point, the effective superpotential encodes large volume disk instantons of the topological A-model mirror geometry \refs{\OoguriBV,\WittenFB,\KachruIH}. We expect that our methods are also applicable to such situations and therefore allow us to determine these integer invariants for D5-branes in suitable compact Calabi-Yau geometries.

From the effective superpotential, which we have computed in the B-model, we have extracted orbifold disk invariants. It would be interesting to directly extract these equivariant invariants on the mirror A-model side by adequate localization techniques. Then the comparison with the topological A-model would also fix the overall normalization ambiguity of our B-model computation.

%%%%%%%%%%%%%%%%%%%%%%
\goodbreak\bigskip\noindent 
{\bf Acknowledgments} 
\medskip\noindent
%%%%%%%%%%%%%%%%%%%%%%
We would like to thank Vincent Bouchard, Bogdan Florea, Shamit Kachru, Johanna Knapp, Wolfgang Lerche, Emanuel Scheidegger, and especially Peter Mayr for fruitful discussions and helpful correspondences. M.~S. is grateful to the Simons Workshop in Mathematics and Physics 2008 for its stimulating atmosphere, where part of this work was completed. This work is supported by the Stanford Institute for Theoretical Physics and by the NSF~Grant~0244728.
%%%%%%%%%%%%%%%%%%%%%%
\goodbreak\bigskip
%%%%%%%%%%%%%%%%%%%%%%

%%%%%%%%%%%%%%%%%%%%%%
\appendix{A}{The degree-eight hypersurface example in ${\bf\IW\IP^{4}_{(1,1,1,1,4)}/(\IZ_8)^2\times\IZ_2}$}
%%%%%%%%%%%%%%%%%%%%%%
In this appendix we give some further details that lead to the expressions for the connection matrices \MPhiOne~and \MPsiOne\ for our first example. The basic idea is to extend the Griffiths-Dwork algorithm to the residue integrals of relative three forms. For ease of notation let us first introduce the abbreviations, ${\uu}\equiv x_{1}x_{2}x_{3}x_{4}x_{5}$ and ${\vv}=x_{1}x_{2}x_{3}x_{4}$. Then together with eqs.~\POne~and \PolDivOne\ we arrive at the simple relation
\eqn\vQdPone{{\vv}={Q\over 4\psi-\phi}-{\partial_{5}P\over 2(4\psi-\phi)}\ .}
In order to approach the problem systematically we need to find relations among all derivatives of the relative holomorphic three form~\holthree~in the filtration~\VarMixed.

Let us first focus on the equations governing the sub-system~\VarSubSys. If we consider polynomial, $Q$, as another constraint besides the defining polynomial equation, $P$, then this defines a complete intersection manifold, and as a result the system of equations for the sub-system has to close by itself. Using the relation~\vQdPone\ and integrating by parts, which corresponds to adding an appropriate exact forms~\dbeta, we find the following relations
\eqn\gamtobetaone{\eqalign{\partial_{\psi}\partial_{\phi}\rel{\Omega}&\simeq{4\phi\over 4\psi-\phi}\,\partial_{\phi}^{2}\rel{\Omega}\ ,\cr
\partial_{\psi}\partial^{2}_{\phi}\rel{\Omega}&\simeq{4\phi\over 4\psi-\phi}\,\partial^{3}_{\phi}\rel{\Omega}+{16\psi\over (4\psi-\phi)^{2}}\,\partial^{2}_{\phi}\rel{\Omega}\ ,\cr
\partial^{2}_{\psi}\partial_{\phi}\rel{\Omega}&\simeq{16\phi^{2}\over (4\psi-\phi)^{2}}\,\partial^{3}_{\phi}\rel{\Omega}+{16\phi^{2}\over (4\psi-\phi)^{3}}\,\partial^{2}_{\phi}\rel{\Omega}\ .}}
If we differentiate the relative three form, $\rel{\Omega}$, once more, then according to the variational diagram~\VarMixed, these derivatives cannot be independent anymore, but instead must be expressible in terms of the lower derivatives up to exact forms. Therefore, using again eqs.~\vQdPone~and~\dbeta, we find the following three relation among fourth-order derivatives
\eqn\gamrel{\eqalign{\partial_{\psi}\partial^{3}_{\phi}\rel{\Omega}&\simeq {4\phi\over 4\psi-\phi}\,\partial^{4}_{\phi}\rel{\Omega}+{32\psi\over (4\psi-\phi)^{2}}\,\partial^{3}_{\phi}\rel{\Omega}+{32\psi\over (4\psi-\phi)^{3}}\,\partial^{2}_{\phi}\rel{\Omega}\ ,\cr
\partial^{2}_{\psi}\partial^{2}_{\phi}\rel{\Omega}&\simeq {4\phi\over 4\psi-\phi}\,\partial_{\psi}\partial^{3}_{\phi}\rel{\Omega}+{16\psi\over (4\psi-\phi)^{2}}\,\partial_{\psi}\partial^{2}_{\phi}\rel{\Omega}-{16\phi\over (4\psi-\phi)^{2}}\,\partial^{3}_{\phi}\rel{\Omega}-{16(4\psi+\phi)\over (4\psi-\phi)^{3}}\,\partial^{2}_{\phi}\rel{\Omega}\ ,\cr
\partial^{3}_{\psi}\partial_{\phi}\rel{\Omega}&\simeq {4\phi\over 4\psi-\phi}\,\partial^{2}_{\psi}\partial^{2}_{\phi}\rel{\Omega}-{32\,\phi\over (4\psi-\phi)^{2}}\,\partial_{\psi}\partial^{2}_{\phi}\rel{\Omega}+{128\,\phi\over (4\psi-\phi)^{3}}\,\partial^{2}_{\phi}\rel{\Omega}\ .}}
However, in order to eventually close the sub-system we need one more non-trivial relation involving the fourth-order derivatives of the relative form, $\rel\Omega$. To get this last equation we first observe the following algebraic equation holds
\eqn\phicloseone{\big(1-(2\psi)^{8}\big){\uu}^{3}{\vv}=I_{1}+I_{2}+I_{3}+I_{4}+I_{5}+I_{6}+I_{7}+I_{8}\ ,}
where polynomials, $I_{j}$, are given by
\eqn\defItwo{\eqalign{I_{1}&={1\over 2}\,{\vv}^{4}x_{5}^{2}\,\partial_{5}P\ ,\qquad\qquad\qquad\qquad\  I_{2}=2\psi\,{\vv}^{5}x_{5}\partial_{5}P\ ,\cr
I_{3}&=8\psi^{2}\,{\vv}^{6}\partial_{5}P\ ,\qquad\qquad\qquad\qquad\ \  I_{4}=8\psi^{3}\,(x_{1}x_{2}x_{3})^{7}\partial_{4}P\ ,\cr
I_{5}&=8\psi^{4}\,(x_{1}x_{2})^{8}x_{3}x_{5}\partial_{3}P\ ,\qquad\qquad\  I_{6}=8\psi^{5}\,x_{1}^{9}x_{2}^{2}x_{3}x_{4}x_{5}^{2}\,\partial_{2}P\ ,\cr
I_{7}&=8\psi^{6}\,x_{1}^{3}(x_{2}x_{3}x_{4})^{2}x_{5}^{3}\,\partial_{1}P\ ,\qquad\quad  I_{8}=32\psi^{7}\,{\uu}^{3}\partial_{5}P\ .}}
If we now use eq.~\dbeta\ repeatedly, after a long and tedious computation we obtain
\eqn\closephione{\eqalign{\big(1-(2\psi)^{8}\big)&\,\partial^{3}_{\psi}\partial_{\phi}\rel{\Omega}\simeq -128 \phi  (\phi -8 \psi ) \psi ^3\partial_{\phi}\rel{\Omega}\cr
&-16\Big({192\psi^{2}\over (4\psi-\phi)^{4}}-{48\psi\over (4\psi-\phi)^{3}}+{8\over (4\psi-\phi)^{2}}+19\phi^{3}\psi^{3}-228\phi^{2}\psi^{4}\Big)\partial^{2}_{\phi}\rel{\Omega}\cr
&-16\psi\Big({192\psi\over (4\psi-\phi)^{3}}-{32\over (4\psi-\phi)^{2}}+9\phi^{4}\psi^{2}-144\phi^{3}\psi^{3}\Big)\partial^{3}_{\phi}\rel{\Omega}\cr
&+1600\phi\psi^{5}\partial_{\psi}\partial_{\phi}\rel{\Omega}-32\Big({2\psi\over (4\psi-\phi)^{2}}-{1\over (4\psi-\phi)}-60\phi^{2}\psi^{5}\Big)\partial_{\psi}\partial^{2}_{\phi}\rel{\Omega}\cr
&+128 \psi ^6 (5 \phi +4 \psi )\partial^{2}_{\psi}\partial_{\phi}\rel{\Omega}-16\psi^{2}\Big({64\over (4\psi-\phi)^{2}}+\phi^{5}\psi-20\phi^{4}\psi^{2}\Big)\partial^{4}_{\phi}\rel{\Omega}\cr
&-64\psi\Big({1\over 4\psi-\phi}-6\phi^{3}\psi^{4}\Big)\partial_{\psi}\partial^{3}_{\phi}\rel{\Omega}-4 \big(64 \phi\psi^{6}(\psi-\phi)+1\big)\partial^{2}_{\psi}\partial^{2}_{\phi}\rel{\Omega}\ .}}
Taking the eqs.~\gamrel\ and \closephione, we can now express all fourth (and all higher order) derivatives of the holomorphic relative three form, $\rel\Omega$, in terms of the lower order derivatives. In particular, if we choose the basis~\BasisOne\ and express all derivatives in terms of this basis, then we exactly arrive at eq.~\DPhiPiFour. We can then proceed and compute in terms of the chosen basis~\BasisOne\ the connection matrix, $M_{\phi}$, given in eq.~\MPhiOne.

So far, we have closed the system of linear differential equations with respect to the open-string modulus, $\phi$. We also need to close the system with respect to the closed-string modulus, $\psi$. As in eqs.~\gamrel\ and \closephione\ this is achieved by expressing $\partial^{4}_{\psi}\rel{\Omega}$ in terms of lower order derivatives. Note that $\partial^{4}_{\psi}\rel{\Omega}$ is the only additional three form we need to consider, as all the other fourth order derivatives appear in eqs.~\gamrel\ and \closephione.

If we multiply eq.~\phicloseone\ by $x_{5}$, then we find
\eqn\ufourone{{\uu}^{4}={1\over \big(1-(2\psi)^{8}\big)}\sum_{j=1}^{8}x_{5}I_{j}\ .}
Similarly to treatment of the variational sub-system, if we apply the relations~\dbeta~and~\dRalpha\ repeatedly in order to reduce the right hand side of eq~\ufourone~to lower degree, we find
\eqn\closedeltaone{\eqalign{\big(1-(2\psi)^{8}\big)&\,\partial^{4}_{\psi}\rel{\Omega}\,\simeq 256\,\psi^{4}\,\rel{\Omega}+{15\over {\psi}^{3}}(1+256\,\psi^{8})\partial_{\psi}\rel{\Omega}-{5\over {\psi}^{2}}(3-1280\,\psi^{8})\partial^{2}_{\psi}\rel{\Omega}\cr
&+{2\over \psi}(3+1280\,{\psi}^{8})\partial^{3}_{\psi}\rel{\Omega}+{1\over {\psi}^{3}}\big(256\, \phi  (2 \phi -\psi ) \psi ^6+60\big)\partial_{\phi}\rel{\Omega}+1216\,\phi^{3}\psi^{3}\partial^{2}_{\phi}\rel{\Omega}\cr
&+576\,\phi^{4}\psi^{3}\partial^{3}_{\phi}\rel{\Omega}+{4\over \psi^{2}\phi}\big(\phi(64 \phi  (8 \phi -7 \psi ) \psi ^6-15)-12\psi\big)\partial_{\psi}\partial_{\phi}\rel{\Omega}\cr
&+1664\,\phi^{3}\psi^{4}\partial_{\psi}\partial^{2}_{\phi}\rel{\Omega}+8\Big({8 \psi\over \phi ^2}+{6\over\phi }+{3\over\psi }-192\, \phi\psi^{5}(\psi-\phi)\Big)\partial^{2}_{\psi}\partial_{\phi}\rel{\Omega}\cr
&+64\,\phi^{5}\psi^{3}\partial^{4}_{\phi}\rel{\Omega}+256\,\phi^{4}\psi^{4}\partial_{\psi}\partial^{3}_{\phi}
\rel{\Omega}+384\,\phi^{3}\psi^{5}\partial^{2}_{\psi}\partial^{2}_{\phi}\rel{\Omega}\cr
&-{4\over \phi^{2}}\Big(\phi^{2}+4\,\phi\,\psi+16\,\psi^{2}+64\,\phi^{3}\psi^{6}(\psi-\phi)\Big)\partial^{3}_{\psi}
\partial_{\phi}\rel{\Omega}\ .}}
Note that if we set the open-string modulus, $\phi$, and its derivatives, $\partial_\phi$, to zero, then we exactly recover the relevant equation for closed-string variation~\VarBulk\ of the holomorphic three form, $\Omega$. Again, if we choose the basis~\BasisOne\ and rewrite everything in terms of this basis, then we arrive at the relation~\DPsiPiThree. Furthermore with eq.~\closedeltaone we can also compute the connection matrix, $M_{\psi}$, presented in eq.~\MPsiOne.

%%%%%%%%%%%%%%%%%%%%%%%%%%%%%%
\appendix{B}{The mirror quintic example}
%%%%%%%%%%%%%%%%%%%%%%%%%%%%%%
In this appendix we provide some additional computational detail for the extended Griffiths-Dwork algorithm applied to our second example, \ie D5-branes in the mirror of the quintic threefold. Let us first introduce introduce the relevant residue integrals for the basis elements~\BasisTwo. They are given by
\eqn\BasisTwo{\eqalign{\pi_{3-k,3}&=k!\,5^{k}\int {{\uu}^{k}\over P(\psi)^{1+k}}\,\log Q(\phi)\Delta\ ,\qquad k\,=\,0,1,2,3 \ ,\cr
\pi_{2-l,4}&=-l!\,5^{l}\int {{\uu}^{l}{\vv}\over P(\psi)^{1+l}Q(\phi)}\,\Delta\ ,\qquad l\,=\,0,1,2\ ,}}
in terms of the polynomials, $P(\psi)$ and $Q(\phi)$, defined in eqs.~\PTwo\ and \PolDivTwo. Analogously to the previous example the monomials, $x_{1}x_{2}x_{3}x_{4}x_{5}$ and $x_{1}x_{2}x_{3}x_{4}$, are abbreviated by $\uu$ and $\vv$ respectively. In order to obtain the Gauss-Manin connection, we need to find the derivatives of the above basis elements with respect to both closed-string and open-string moduli. As in the other example, some of the derivatives are trivial and, in particular, the relations~\TrivRelations\ are obviously also valid here. To find the other relations, we first realize that
\eqn\vQdPtwo{\vv={Q\over \psi-\phi}-{\partial_{5}P\over 5(\psi-\phi)}\ .}
and together with eq.~\dbeta, we arrive at
\eqn\gamtobetatwo{\eqalign{\partial_{\psi}\partial_{\phi}\rel{\Omega}&\simeq{4\phi\over \psi-\phi}\,\partial^{2}_{\phi}\rel{\Omega}+{3\over \psi-\phi}\,\partial_{\phi}\rel{\Omega}\ ,\cr
\partial_{\psi}\partial^{2}_{\phi}\rel{\Omega}&\simeq{4\phi\over \psi-\phi}\,\partial^{3}_{\phi}\rel{\Omega}+{7\psi-3\phi\over (\psi-\phi)^{2}}\,\partial^{2}_{\phi}\rel{\Omega}+{3\over (\psi-\phi)^{2}}\,\partial_{\phi}\rel{\Omega}\ ,\cr
\partial^{2}_{\psi}\partial_{\phi}\rel{\Omega}&\simeq{16\phi^{2}\over (\psi-\phi)^{2}}\,\partial^{3}_{\phi}\rel{\Omega}+{4\phi(9\psi-5\phi)\over (\psi-\phi)^{3}}\,\partial^{2}_{\phi}\rel{\Omega}
+{6(\psi+\phi)\over (\psi-\phi)^{3}}\,\partial_{\phi}\rel{\Omega}\ .}}
With these relations we express on the level of cohomology all two forms in terms of the chosen basis elements $\{\pi_{2,4},\pi_{1,4},\pi_{0,4}\}$. That is to say, if we now differentiate any two form cohomology element one more time with respect to either one of the two moduli the result is entirely expressible in terms of mentioned basis elements. Similarly to the previous example, with eqs.~\vQdPtwo\ and~\dbeta\ we obtain the following relations
\eqn\gamreltwo{\eqalign{\partial_{\psi}\partial^{3}_{\phi}\rel{\Omega}&\simeq{4\phi\over \psi-\phi}\,\partial^{4}_{\phi}\rel{\Omega}+{11\over \psi-\phi}\,\partial^{3}_{\phi}\rel{\Omega}+{2\over \psi-\phi}\,\partial_{\psi}\partial^{2}_{\phi}\rel{\Omega}\ ,\cr
\partial^{2}_{\psi}\partial^{2}_{\phi}\rel{\Omega}&\simeq{4\phi\over \psi-\phi}\,\partial_{\psi}\partial^{3}_{\phi}\rel{\Omega}+{1\over \psi-\phi}\,\partial^{2}_{\psi}\partial_{\phi}\rel{\Omega}+{6\over \psi-\phi}\,\partial_{\psi}\partial^{2}_{\phi}\rel{\Omega}\ ,\cr
\partial^{3}_{\psi}\partial_{\phi}\rel{\Omega}&\simeq{4\phi\over \psi-\phi}\,\partial^{2}_{\psi}\partial^{2}_{\phi}\rel{\Omega}+{1\over \psi-\phi}\,\partial^{2}_{\psi}\partial_{\phi}\rel{\Omega}\ .}}
To close the system of linear differential equations with respect to the open-string modulus, $\phi$, we need one more equation. The last non-trivial equation is obtained by observing that
\eqn\phiclosetwo{(1-\psi^{5}){\uu}^{3}\vv=I_{1}+I_{2}+I_{3}+I_{4}+I_{5}\ ,}
where the polynomials, $I_{j}$, are given by
\eqn\defI{\eqalign{I_{1}&= {1\over 5}\,(x_{2}x_{3}x_{4})^{4}x_{5}^{3}\,\partial_{1}P\ ,\cr
                   I_{2}&= {1\over 5}\,\psi\,x_{2}(x_{3}x_{4})^{5}x_{5}^{4}\,\partial_{2}P\ ,\cr
                   I_{3}&= {1\over 5}\,\psi^{2}\,x_{1}x_{2}x_{3}^{2}x_{4}^{6}x_{5}^{5}\,\partial_{3}P\ ,\cr
                   I_{4}&= {1\over 5}\,\psi^{3}\,(x_{1}x_{2}x_{3})^{2}x_{4}^{3}x_{5}^{6}\,\partial_{4}P\ ,\cr
                   I_{5}&= {1\over 5}\,\psi^{4}\,{\uu}^{3}\,\partial_{5}P\ .}}
Thus with eq.~\phiclosetwo\ we find for the derivative, $\partial^{3}_{\psi}\partial_{\phi}\rel{\Omega}$,
\eqn\closephiEqtwo{\eqalign{(1-&\,\psi^{5})\,\partial^{3}_{\psi}\partial_{\phi}\rel{\Omega}\simeq-2\phi\,(\phi-5\psi)
\partial_{\phi}\rel{\Omega}-{19\over 4}\phi^{2}(\phi-5\psi)\partial^{2}_{\phi}\rel{\Omega}-{9\over 4}\phi^{3}(\phi-5\psi)\partial^{3}_{\phi}\rel{\Omega}\cr
&-{19\over 4}\psi\phi(\phi-5\psi)\partial_{\psi}\partial_{\phi}\rel{\Omega}-{9\over 2}\psi\phi^{2}(\phi-5\psi)\partial_{\psi}\partial^{2}_{\phi}\rel{\Omega}-{1\over 4}\psi^{2}\big(9\phi(\phi-5\psi)+4\psi^{2}\big)\partial^{2}_{\psi}\partial_{\phi}\rel{\Omega}\cr
&-{1\over 4}\phi^{4}(\phi-5\psi)\partial^{4}_{\phi}\rel{\Omega}-{3\over 4}\psi\phi^{3}(\phi-5\psi)\partial_{\psi}\partial^{3}_{\phi}\rel{\Omega}-{1\over 4}\psi^{2}\phi\big(3\phi(\phi-5\psi)+16\psi^{2}\big)\partial^{2}_{\psi}\partial^{2}_{\phi}\rel{\Omega}\cr
&-{1\over 4}\psi^{3}\phi(\phi-\psi)\partial^{3}_{\psi}\partial_{\phi}\rel{\Omega}\ .}}
We should also need to close the linear system with respect to the closed-string modulus, $\psi$. To determine this last relevant equation, we first note that the following algebraic relation holds
\eqn\ufourtwo{\eqalign{\Big(1-{1\over {\psi}^{5}}\Big){\uu}^{4}=&-{1\over 5\psi}{\uu}^{3}x_{1}\partial_{1}P-{1\over 5\psi}{\uu}^{2}x_{1}^{5}x_{2}\partial_{2}P
-{1\over 5\psi}{\uu}\,(x_{1}x_{2})^{5}x_{3}\partial_{3}P\cr
&-{1\over 5\psi}(x_{1}x_{2}x_{3})^{5}x_{4}\partial_{4}P-{1\over 5\psi}{\vv}^{4}\partial_{5}P\ .}}
We now use eqs.~\dbeta\ and \dRalpha\ to express all the residue integrals resulting from the right  hand side of eq.~\ufourtwo\ in terms of derivatives of the holomorphic relative three form, $\rel\Omega$. After a long computation we obtain
\eqn\closedeltatwo{\eqalign{\partial^{4}_{\psi}\rel{\Omega}\simeq{\psi^{5}\over 1-\psi^{5}}\Big(&\psi\,\rel{\Omega}+15\psi^{2}\partial_{\psi}\rel{\Omega}+25\psi^{3}\partial^{2}_{\psi}\rel{\Omega}
+10\psi^{4}\partial^{3}_{\psi}\rel{\Omega}+15\psi\phi\,\partial_{\phi}\rel{\Omega}+25\psi\phi^{2}
\partial^{2}_{\phi}\rel{\Omega}\cr
&+10\psi\phi^{3}\partial^{3}_{\phi}\rel{\Omega}+50\phi\psi^{2}
\partial_{\psi}\partial_{\phi}\rel{\Omega}+30\phi^{2}\psi^{2}\partial_{\psi}\partial^{2}_{\phi}\rel{\Omega}
+30\phi\psi^{3}\partial^{2}_{\psi}\partial_{\phi}\rel{\Omega}\cr
&+\phi^{4}\psi\partial^{4}_{\phi}\rel{\Omega}+4\phi^{3}\psi^{2}
\partial_{\psi}\partial^{3}_{\phi}\rel{\Omega}+6\phi^{2}\psi^{3}\partial^{2}_{\psi}
\partial^{2}_{\phi}\rel{\Omega}+4(\phi\psi^{4}-1)\partial^{3}_{\psi}\partial_{\phi}\rel{\Omega}\,\Big)\ .}}
Now we have a complete system of linear differential equations at hand, which allows us to determine the needed linear combinations of differentiated basis elements with respect to both open- and closed-string moduli. The non-trivial derivatives of the two-form basis elements are given by
\eqn\DPhiPiFourtwo{\eqalign{
  \partial_\phi\pi_{2,4}\,\simeq&\, {\psi-\phi\over 4\phi}\,\pi_{1,4}-{3\over 4\phi}\,\pi_{2,4}  \ , \cr
  \partial_\phi\pi_{1,4}\,\simeq&\, {\psi-\phi \over 4\phi}\,\pi_{0,4}-{1\over 2\phi}\,\pi_{1,4} \ , \cr
  \partial_\phi\pi_{0,4}\,\simeq&\, -{125(\phi-\psi)(\phi-5\psi)\over 4D_{2}}\,\pi_{2,4}+{175(\phi-\psi)(\phi-5\psi)^{2}\over 4D_{2}}\,\pi_{1,4}\cr
  &\,-\Big({1\over 4\phi}+{15(\phi-\psi)(\phi-5\psi)^{3}\over 2D_{2}}\Big)\,\pi_{0,4}\ ,\cr
  \partial_{\psi}\pi_{0,4}\,\simeq&\, {125\,\phi(\phi-5\psi)\over D_{2}}\,\pi_{2,4}-{175\phi(\phi-5\psi)^{2}\over D_{2}}\,\pi_{1,4}+{30\,\phi(\phi-5\psi)^{3}\over D_{2}}\,\pi_{0,4}\ .}}
There are also two non-trivial derivative relations in the three-form sector, namely the derivatives of the basis element, $\pi_{0,3}$, with respect to both the open- and closed-string moduli. After going through another long calculation, we find the relations
\eqn\dPhiThreetwo{\partial_{\phi}\pi_{0,3}\simeq\, {125\,\phi(\phi-5\psi)\over D_{2}}\,\pi_{2,4}-{175\phi(\phi-5\psi)^{2}\over D_{2}}\,\pi_{1,4}+{30\,\phi(\phi-5\psi)^{3}\over D_{2}}\,\pi_{0,4}\ ,}
and
\eqn\DPsiThreetwo{\eqalign{\partial_{\psi}\pi_{0,3}\simeq&\,{\psi\over D_{1}}\,\pi_{3,3}+{15\psi^{2}\over D_{1}}\,\pi_{2,3}+{25\psi^{3}\over D_{1}}\,\pi_{1,3}+{10\psi^{4}\over D_{1}}\,\pi_{0,3}\cr
&-{\phi\, T_{1}\over 16D_{1}\,D_{2}}\,\pi_{2,4}-{\phi\, T_{2}\over 16D_{1}\,D_{2}}\,\pi_{1,4}-{\phi\, T_{3}\over 16D_{1}\,D_{2}}\,\pi_{0,4}\ ,}}
Here the polynomials $T_{1}$, $T_{2}$, $T_{3}$, and the discriminants, $D_1$ and $D_2$, are respectively defined in eqs.~\threeT\ and \DiscQuintic.

With~eqs.~\DPhiPiFourtwo, \dPhiThreetwo, and \DPsiThreetwo, we can now easily extract the connection matrices, $M_{\phi}$ and $M_{\psi}$, which were presented in \MPsiTwo\ and \MPhiTwo.

%%%%%%%%%%%%%%%%%%%%%%
\listrefs
\end